\documentclass[11pt, dvips]{article}

\RequirePackage[OT1]{fontenc}
\usepackage{amsthm,amsmath,amssymb}
\usepackage{graphicx}
\RequirePackage[numbers]{natbib}
\RequirePackage[dvips, colorlinks]{hyperref}

\RequirePackage{hypernat}

\def\eqdef{ \mbox{\raisebox{-0.4ex}{$\stackrel{\rm def}=$}} }
\def\timess{ {\kern-0.1ex\times\kern-0.1ex} }

\font\eufrak=eufm10
\newcommand{\gotL}{\mbox{\eufrak{L}}}
\newcommand{\gotR}{\mbox{\eufrak{R}}}

\newcommand{\bbC}{\mathbb{C}}
\newcommand{\bbR}{\mathbb{R}}
\newcommand{\bbN}{\mathbb{N}}
\newcommand{\bbZ}{\mathbb{Z}}
\newcommand{\Cdisk}{\mathbb{D}}
\newcommand{\Ccircle}{\mathbb{T}}

\def\calR{\mathcal{R}}
\def\calB{\mathcal{B}}
\def\calA{\mathcal{A}}
\def\calD{\mathcal{D}}
\def\calN{\mathcal{N}}

\def\bfP{\textbf{P}}
\def\bfL{\textbf{L}}
\def\bfT{\textbf{T}}
\def\bfX{\textbf{X}}
\def\rmT{\texttt{T}}
\def\rmR{\textrm{R}}

\newtheorem{theorem}{Theorem}[section]
\newtheorem{lemma}{Lemma}[section]
\newtheorem{proposition}{Proposition}[section]
\newtheorem{corollary}{Corollary}[section]
\newtheorem{definition}{Definition}[section]
\newtheorem{example}{Example}[section]
\newtheorem{algorithm}{Algorithm}[section]
\newtheorem{remark}{Remark}[section]

\newcommand{\bt}{\begin{theorem}}
\newcommand{\et}{\end{theorem}}
\newcommand{\bl}{\begin{lemma}}
\newcommand{\el}{\end{lemma}}
\newcommand{\bp}{\begin{proposition}}
\newcommand{\ep}{\end{proposition}}
\newcommand{\bc}{\begin{corollary}}
\newcommand{\ec}{\end{corollary}}
\newcommand{\btb}{\begin{table}[hbp]}
\newcommand{\btbh}{\begin{table}[!ht]}
\newcommand{\etb}{\end{table}}
\newcommand{\bfg}{\begin{figure}[!ht]}
\newcommand{\bfgt}{\begin{figure}[t]}
\newcommand{\bfgb}{\begin{figure}[b]}
\newcommand{\bfgh}{\begin{figure}[!ht]}
\newcommand{\efg}{\end{figure}}
\newcommand{\bd}{\begin{definition}\rm}
\newcommand{\ed}{\end{definition}}
\newcommand{\bex}{\begin{example}\rm}
\newcommand{\eex}{\end{example}}
\newcommand{\ba}{\begin{algorithm}\rm}
\newcommand{\ea}{\end{algorithm}}
\newcommand{\br}{\begin{remark}\rm}
\newcommand{\er}{\end{remark}}
\newcommand{\bea}{\begin{eqnarray*}}
\newcommand{\eea}{\end{eqnarray*}}
\newcommand{\be}{\begin{eqnarray}}
\newcommand{\ee}{\end{eqnarray}}
\newcommand{\suml}{\sum\limits}
\newcommand{\intl}{\int\limits}

\newcommand{\ovl}{\overline}
\newcommand{\lm}{\lambda}

\def\span{\mathop{\rm span}}
\def\iim{\mathop{\rm Im}}

\def\what{\widehat}

\newcommand{\up}{{\triangledown}}

\begin{document}


\title{On signal and extraneous roots in Singular Spectrum Analysis}

\author{Konstantin Usevich \\
Department of Mathematics and Mechanics, \\
St.Petersburg State University, Russia \\
E-mail: \url{konstantin.usevich@statmod.ru}}


\maketitle

\begin{abstract}
In the present paper we study properties of roots of 
characteristic polynomials for the linear recurrent formulae (LRF) that
govern time series. We also investigate how the values of these roots affect Singular Spectrum Analysis
implications, in what concerns separation of components, SSA forecasting and related signal parameter estimation methods.
 The roots of the characteristic polynomial for an LRF comprise 
the signal roots, which determine the structure of the time series, and extraneous roots. 
We show how the separability  of two time series can be characterized in terms of their signal roots. 
All possible cases of exact separability are enumerated.
We also examine properties of extraneous roots of the LRF used in SSA forecasting  algorithms, which is equivalent to the Min-Norm vector in subspace-based estimation methods. We apply recent theoretical results for orthogonal polynomials on the unit circle, which enable us 
to precisely describe the asymptotic distribution of extraneous roots relative to the position of the signal roots. 
\end{abstract}

\noindent{\bf Keywords:}
Singular Spectrum Analysis; SSA; separability; linear recurrent formula; LRF;
continuation; extraneous roots; min-norm; subspace methods;
orthogonal polynomials on the unit circle;

%





\section{Introduction}
In the theory of Singular Spectrum Analysis \cite{ET1996,GNZH2001} the time series 
that are governed by a \textit{linear recurrent formula} (LRF) are of great importance.
For these time series $F_N = (f_0,\ldots,f_{N-1})$  there exist coefficients $a_k$ such that the relation
\be\label{eq:LRF_INTRO}
f_{\rho +n} = \suml_{k=0}^{\rho-1} a_k f_{k+n},  
\ee
$a_0 \neq 0$, holds for all appropriate $n$; these time series are called the \textit{time series of finite difference dimension}.
In the present paper we consider complex-valued time series and LRFs with complex coefficients.

The roots of the \textit{characteristic polynomial}   $A(z) \eqdef z^{\rho}-a_{\rho-1} z^{\rho-1} + \ldots + a_1 z+ a_0$
of the LRF \eqref{eq:LRF_INTRO} can be divided into two groups:
 the \textit{signal roots} which determine the representation of the time series
as a sum of polynomially modulated exponential signals
\be\label{eq:PME_INTRO}
f_n = \suml_{k=1}^{m} P_{k}(n) \lm_k^n,
\ee
and other roots, called \textit{extraneous roots}. In this paper we show how the signal roots of the LRFs determine several 
properties of time series, in what concerns the Singular Spectrum Analysis.
Specifically, we consider separability and the behaviour of extraneous roots for the LRF used in SSA forecasting. The latter LRF is of particular interest since the extraneous roots affect the results of SSA continuation.
The obtained results also concern \textit{subspace-based} methods of signal processing \cite{KT1982,KFP1992,SM1997,BDR2006} 
(in particular, the \textit{Min-Norm} method), which share their structure with SSA.
 
The first part of this paper is devoted to a novel outlook at separability \cite[Ch. 6]{GNZH2001}.
A new criterion of exact weak separability for the time series of finite difference dimension is introduced. This criterion enables us to present all the standard examples  in a unified manner, elucidating the meaning of the conditions of separability in these examples. Moreover, it enables us to enumerate all possible cases of  exact weak separability for arbitrary time series; this is also a new result. In fact, we develop a criterion for the one-sided separability, which has been recently shown to be important for the SSA theory \cite{N2010}.  The criterion  is simple, illustrative and based solely on the signal roots of the time series. Specific features of the case
where the time series is real are clarified as well.

The second part of the paper deals with the LRF used in SSA forecasting (shortly, the SSA LRF), see \cite[Ch. 5]{GNZH2001}. This LRF coincides with the \textit{Min-Norm prediction vector} \cite{KT1982, KFP1992, SM1997}. Properties of the extraneous roots of the SSA LRF are examined in the noise-free case. Recently, the basic properties were proved by various authors. In particular, it was proved that all extraneous roots lie inside the unit circle and that the extraneous roots of the forward and backward SSA LRF coincide, see \cite{K1983,BB1998}. The asymptotic distribution of the extraneous roots was independently addressed in \cite{N1999, P1987}, 
but only some particular cases were considered. In the present paper we show the correspondence between the extraneous roots of the SSA LRF and 
\textit{orthogonal polynomials on the unit circle} \cite{S2005}. This correspondence enables us to provide elegant and short proofs for the basic properties of the extraneous roots and to describe the asymptotic distribution of the roots. By the way, this approach was used in \cite{P1987}, but in the present paper we try to provide a more accurate, comprehensive and integral exposition. We describe the asymptotic distribution for the general case of time series of finite difference dimension with the help of the most recent advances in the theory of orthogonal polynomials \cite{MMS2006a,MMS2006s}.

Both parts of the present paper are based on the same ground of a revised theory of  time series of finite difference dimension.
In fact, this revision can be useful in the SSA theory on its own. Surprisingly, the two parts of the present paper 
are linked from another side: the approximate separability of certain time series can be somehow described 
with the help of the distribution of extraneous roots of a specific SSA LRF.

This paper is organized as follows. In Section~\ref{sect:POL_EXP} we provide all necessary background on the
linear recurrent formulae and continuation, the time series of finite difference dimension and their trajectory spaces.
 Section~\ref{sect:SEP} is devoted to exact weak separability.  In Section~\ref{sect:SSA_LRF} we investigate the SSA  
forecasting LRF and the properties of its extraneous roots. At the end of Section~\ref{sect:SSA_LRF} we discuss the connection between the
asymptotic distribution of extraneous roots and separability; we also make practical conclusions and examine the behaviour of roots in the presence of noise as well.

\section{Basic facts and notation}\label{sect:POL_EXP}
\subsection{Time series of finite difference dimension and signal roots}
Let $\bbN_0$ denote the set of all nonnegative integers. An \textit{infinite time series}
\be\label{eq:INF_TS}
F_\infty = (f_0,f_1, \ldots), \quad f_n \in \bbC,
\ee
is said to \textit{satisfy a linear recurrent formula}  (\textit{LRF}) of order $\rho$ if
there exist coefficients $a_0,\ldots,a_{\rho-1} \in \bbC$ such that the relation
\be\label{eq:LRF1}
f_{\rho+n} = \suml_{k=0}^{\rho-1} a_k f_{k+n} 
\ee 
holds for all $n \in \bbN_0$. Note that in the case $\rho = 0$ we have $f_n = 0$ for all $n \in \bbN_0$.
Once a time series satisfies an LRF \eqref{eq:LRF1}, its form can be described by the roots of the \textit{characteristic polynomial} of the LRF
\be\label{eq:LRF_CHAR_POLY}
A(z) = z^\rho - a_{\rho-1} z^{\rho-1} - \ldots - a_1 \rho -a_0.
\ee

\bt[{\cite[Th. 3.1.1]{H1998}}]\label{th:POL_EXP_FROM_LRF}
Assume that an infinite time series $F_\infty$ satisfies an LRF \eqref{eq:LRF1} with $a_0 \neq 0$.
Then it can be represented as
\be\label{eq:pol_exp_sign}
f_n = \suml_{k=1}^{m} P_{k}(n) \lm_k^n, 
\ee
where $\lm_k \in \bbC \setminus \{0\}$ are distinct numbers, and $P_k$ are non-zero polynomials.
All $\lm_k$ in the representation \eqref{eq:pol_exp_sign} are roots of the characteristic polynomial $A(z)$, 
with multiplicity not less than $\nu_k \eqdef \deg P_k +1$, where $\deg \cdot$ is the degree of a polynomial.

The coefficients of $P_k$ are determined by the first $d$ values of the time series, where $d$ is defined as
\be\label{eq:POL_EXP_RANK}
d = \nu_1 + \ldots + \nu_m \le \rho.
\ee
\et

\br\label{rem:POL_EXP_UNIQUE}
If a time series admits a representation of type \eqref{eq:pol_exp_sign}, then this representation is unique.
This follows from the linear independence of the time series of type $g_n = n^k \lm^n$ 
for different $\lm \in \bbC \setminus \{0\}$ and $k \in \bbN_0$.
\er

For a time series of type \eqref{eq:pol_exp_sign}, by Remark~\ref{rem:POL_EXP_UNIQUE}, one can
unambiguously define the polynomial
\be\label{eq:CHAR_POLY}
\begin{array}{c}
P(z) \eqdef (z-\lm_1)^{\nu_1} \cdot \ldots \cdot (z-\lm_m)^{\nu_m} = \\
 = p_d z^d + \ldots + p_1 z + p_0, 
\end{array}
\ee
where $p_d = 1$. This polynomial is called the \textit{characteristic polynomial of the time series}.
The characteristic polynomial determines the set of all LRF that are satisfied by the time series.

\bt\label{th:LRF_FROM_P}
Let $F_\infty$ be a time series of the form \eqref{eq:pol_exp_sign}.
Then any polynomial
\be\label{eq:LRF_FROM_VEC}
B(z) = b_r z^r + \ldots b_1 z + b_0
\ee
of degree $r$ (i.e. $b_r \neq 0$) is a multiple of the characteristic polynomial 
\eqref{eq:CHAR_POLY}, i.e. $B(z) = P(z) Q(z)$, if and only if 
the time series satisfies the LRF
\be\label{eq:LRF2}
f_{n+r} = \suml_{k=0}^{r-1} -\frac{b_k}{b_r} f_{n+k}, \quad n \in \bbN_0.
\ee
\et
\begin{proof}
$\boxed{\Rightarrow}$
If $F$ satisfies an LRF $f_{r+n} = \suml_{k=0}^{r-1} b_k f_{k+n}$, which has the characteristic polynomial
$B(z) = z^r - b_{r-1} z^{r-1} - \ldots - b_0$,
then it satisfies the LRF $f_{(r+l)+n} = \suml_{k=l}^{(r+l)-1} b_k f_{k+n} = 0$, which corresponds to $B(z) z^l$. Therefore,
for any $A(z) = B(z) Q(z)$, $F$ satisfies the corresponding LRF.

The proof that a time series of the form \eqref{eq:pol_exp_sign} satisfies the LRF with the characteristic polynomial $P(z)$ can be found in the proof of \cite[Th. 3.1.1]{H1998}.

$\boxed{\Leftarrow}$ Note that if $F_\infty$ satisfies the LRF with the characteristic polynomial $Q(z) = S(z) z^m$, $S(0) \neq 0$, then
the time series $G_\infty = (f_m, f_{m+1}, \ldots)$ satisfies the LRF corresponding to $S(z)$. 
Let $R(z) = \textbf{GCD} (P(z),S(z))$ be the greatest common divisor of $P(z)$ and $S(z)$. 
Then $R(z)$ can be represented in the form $R(z) = P(z) c(z) + S(z) d(z)$ (see \cite[Ch.III, \S17]{W1994}), and
by $\boxed{\Rightarrow}$ part, $G_\infty$ satisfies the LRF with characteristic polynomial  $R(z)$. 
If $R(z)$ is not $P(z)$ then $R(z) =  (z-\lm_1)^{d_1} \cdot \ldots \cdot (z-\lm_m)^{d_m}$ with  
$d_k \le \nu_k$ for all $k$ and at least one $d_l < \nu_l$.
Then by Theorem~\ref{th:POL_EXP_FROM_LRF} $g_n = \suml_{k=1}^{m} Q_{k}(n) \lm_k^n$,
where $Q_l(n)$ has the degree less than $d_k-1$. By the linear independence of time series $n^k \lm^n$ 
we obtain the contradiction with the representation \eqref{eq:pol_exp_sign}. 
\end{proof}

\br
Theorems~\ref{th:POL_EXP_FROM_LRF}~and~\ref{th:LRF_FROM_P} establish the one-to-one correspondence between the time series
of type \eqref{eq:pol_exp_sign} and the time series satisfying at least one LRF \eqref{eq:LRF1} with non-zero last coefficient ($a_0 \neq 0$). 
\er

Now assume that a time series $F_\infty$ satisfies an LRF \eqref{eq:LRF1} with $a_0 \neq 0$. By Theorem~\ref{th:POL_EXP_FROM_LRF}
it has the representation \eqref{eq:pol_exp_sign} and the characteristic polynomial \eqref{eq:CHAR_POLY} is uniquely determined.
By Theorem~\ref{th:LRF_FROM_P}, the relation
\be\label{eq:LRF_FACT}
A(z) = P(z) V(z)
\ee
holds. Moreover, the time series  satisfies all LRFs with characteristic polynomials
of form $B(z) = P(z) Q(z)$, and hence the polynomial $V(z)$ \eqref{eq:LRF_FACT} (and its roots) has no effect on the form of 
the time series. Thus, the $\rho$ roots of the characteristic polynomial $A(z)$ can be divided into two groups:
\begin{enumerate}
\item the $d$ \textit{signal roots} (i.e. the roots of $P(z)$), which determine the structure of the time series,
\item the $\rho-d$ \textit{extraneous roots},
\end{enumerate}
where $d$ is defined in \eqref{eq:POL_EXP_RANK}.
We also say that the signal roots $\lm_k$ of $A(z)$ are the \textit{signal roots of the time series} and $\nu_k$ are their \textit{multiplicities}.

The number $d$ of signal roots has an important interpretation in terms of LRFs; 
this interpretation follows from Theorem~\ref{th:LRF_FROM_P}.

\bc\label{cor:CHP}
If a time series $F_\infty$ satisfies an LRF  \eqref{eq:LRF1} with $a_0 \neq 0$, then the LRF corresponding to the characteristic polynomial  
\eqref{eq:CHAR_POLY} of $F_\infty$
\be\label{eq:PME_MIN_LRF}
f_{d+n} = -\suml_{k=0}^{d-1} p_k f_{k+n}
\ee
has the minimal order $d$ among all LRFs satisfied by  $F_\infty$.
\ec

Note that Corollary~\ref{cor:CHP}  is a characterization of $P(z)$, and can be taken as an alternative definition of the characteristic
polynomial $P(z)$. It also validates the following notation.
\bd
We say that $F_\infty$ is a time series  of finite difference dimension (an \textit{f.d.d. time series}) 
if it satisfies at least one LRF \eqref{eq:LRF1} with $a_0\neq0$. The degree $d$ 
 of the characteristic polynomial, defined in \eqref{eq:POL_EXP_RANK}, is called the \textit{difference dimension} of $F_\infty$.
\ed

\br
The case of LRFs with $a_0 = 0$ can be considered within the same framework along with generalizations of Theorems~\ref{th:POL_EXP_FROM_LRF}~and~\ref{th:LRF_FROM_P}, but we omit this consideration.
\er
For clarity, let us consider a real-valued time series $F_\infty$ of finite difference dimension. 
By Theorem~\ref{th:POL_EXP_FROM_LRF}, it has the form
\be\label{eq:PME_REAL}
&&f_n =  \suml_{l=1}^{s'} P_{l} (n)\, \rho_l^n  + \\\nonumber
&&\phantom{f_n}    +       \suml_{l=s'+1}^{s} P_{l} (n)\, \rho_l^n \cos(2\pi\omega_l n + \varphi_l),
\ee
where $\omega_l$, $\rho_l$ are distinct, $|\omega_l| < 0.5$ and $P_l$ are real polynomials of degree $n_l - 1$.
If we denote $\lm_{l} = \rho_l$, $\nu_l = r_l$ for $l \le s'$ and
$\lm_{2l-s'-1} = \rho_l e^{2\pi i \omega_l}$, $\lm_{2l-s} = \rho_l e^{-2\pi i \omega_l}$, $\nu_{2l-s'-1} = \nu_{2l-s} = n_l$
for $s' < l \le s$, then we obtain the representation \eqref{eq:pol_exp_sign} with $m = 2s -s'$ modulated exponents.

\subsection{Hankel matrices and trajectory spaces}

Let
\be\label{eq:time_series_fin}
F = F_N = (f_0, \ldots, f_{N-1})^\rmT \in \bbC^{N}
\ee
be a (finite) \textit{time series}. The {\em Hankel matrix} generated by the time series (or the {\em trajectory matrix})
is the matrix
\begin{equation}
    \bfX^{(L)}(F_N) \eqdef \left(
    \begin{array}{lllll}
        f_0&f_1&f_2&\ldots&f_{K-1}\\
        f_1&f_2&f_3&\ldots&f_{K}\\
        f_2&f_3&f_4&\ldots&f_{K+1}\\
        \vdots&\vdots&\vdots&\ddots&\vdots\\
        f_{L-1}&f_{L}&f_{L+1}&\ldots&f_{N-1}\\
    \end{array}
\right),
\end{equation}
where the parameter $L$ is called the \textit{window length}, $1 < L < N$ and $K = N-L+1$.

\bd\label{def:FDIM_FINITE}
If $F_N$ is a subseries of  an infinite time series \eqref{eq:INF_TS} of difference dimension  $d \le N/2$, then $F_N$ is called a time series \textit{of (finite) difference dimension} $d$ (with characteristic polynomial $P(z)$).
\ed
This definition agrees with the definition given in \cite[Ch.~2]{GNZH2001}, see remarks at the beginning of Section~\ref{ssect:LRF_CONT}.
In particular, the following theorem states that the time series of finite difference dimension are \textit{time series of finite rank}, 
see also \cite[Ch.~5, Prop.~5.4]{GNZH2001}.
\bp\label{prop:POL_EXP_FIN_RANK}
Let $F_N$  be of difference dimension  $d$.
\begin{enumerate}
\item For the window length $L$ such that $d \le L \le N-d+1$
the trajectory matrix \eqref{eq:time_series_fin} is of rank $d$. 
\item If $L < d$ or $L > N-d+ 1$ then
{\rm $\bfX^{(L)}$} has maximal possible rank ($L$ or $N-L+1$, respectively).
\end{enumerate}
\ep
The proposition immediately follows from \cite[Ch. XVI, \S 10, Th. 7]{G1998} and its corollary.

Let us show how the structure of a time series is connected to LRFs which are satisfied by the time series.
The structure of a time series in SSA is described by its \textit{trajectory space}
\bea
\gotL^{(L)} = \gotL^{(L)}(F_N) \eqdef \span(X^{(L)}_1,\ldots,X^{(L)}_{K}) \subseteq \bbC^{L},
\eea
where
\be\label{eq:lag_vectors}
X^{(L)}_i=(f_{i-1},\ldots,f_{i+L-2})^\rmT, \, 1 \leq i\leq K,
\ee
are the columns of the matrix $\bfX^{(L)} = \bfX^{(L)}(F_N)$, see \cite[Ch. 1]{GNZH2001} for a detailed discussion. 
In what follows, the following subspace of $\bbC^L$ is very useful.
\bd\label{def:REL_SPACE}
The \textit{relations space} is defined as
\bea
\gotR^{(L)} = \gotR^{(L)}(F_N) \eqdef \ovl{\gotL^{(L)}_{\bot}},
\eea 
where $\ovl{\gotL}$ denotes the complex conjugation of $\gotL$
and $\gotL^{(L)}_{\bot}$ is the orthogonal complement to the trajectory space.
\ed
The relations space consists of all linear relations on rows of $\bfX^{(L)}$.
Indeed, the vector 
\bea
(a_0,\ldots,a_r, 0, \ldots,0)^\rmT,\quad a_r \neq 0,
\eea
 belongs to $\gotR^{(L)}$ if and only if
\be\label{eq:LRF_FROM_ORTH}
f_{n+r} = -\suml_{k=0}^{r-1} \frac{a_k}{a_r} f_{n+k},\; 0 \le n \le N - L + 1.
\ee
\br\label{ref:ORTH_CONJ_REAL}
In complex vector spaces the inner product  involves complex conjugation. This explains the presence of the conjugation in the definition of the relations space.
 If $F_N$ is real-valued, then $\gotR^{(L)} = \gotL^{(L)}_{\bot}$.
\er

The following proposition shows that the relations space of a time series of finite difference dimension
 is generated by all LRF of order less than $L$, satisfied by its infinite time series (c.f. Definition~\ref{def:FDIM_FINITE}). 

\bp\label{prop:ORTH_TRAJ_REPR}
Let $F_N$ be a time series of difference dimension $d$ with characteristic polynomial $P(z)$ \eqref{eq:CHAR_POLY}.
For the window length $L$, $d < L \le N-d+1$,  we have the following.
\begin{enumerate}
\item The columns of the $L\timess (L-d)$ matrix
{\rm
\be\label{eq:MULT}
\bfP = \bfP^{(L)} = 
\left(
    \begin{array}{llll}
        p_0     & 0      & \ldots &0        \\
        \vdots  & p_0    & \ddots &\vdots   \\
        p_{d}   & \vdots & \ddots &0        \\
        0       & p_{d}  & \ddots &p_0      \\
        \vdots  &\ddots  & \ddots &\vdots   \\
        0       & \ldots & 0      &p_{d}    \\
    \end{array}
\right)
\ee
}
form a basis of the space $\gotR^{(L)}$.
\item
A vector {\rm $B = (b_0,\ldots,b_{L-1})^\rmT$} belongs to $\gotR^{(L)}$ if and only if
$B(z) = b_{L-1} z^{L-1} + \ldots + b_1 z + b_0$ is a multiple of $P(z)$.
\end{enumerate}
\ep
\begin{proof}
Since $p_d \neq 0$, the columns of the matrix $\bfP$ \eqref{eq:MULT} are linearly independent.
By Theorem~\ref{th:LRF_FROM_P}, the time series $F_N$ satisfies the LRF \eqref{eq:PME_MIN_LRF} and the $L-d$ columns of 
 $\bfP$ belong to $\gotR^{(L)}$. By Proposition~\ref{prop:POL_EXP_FIN_RANK},
the dimension of $\gotR^{(L)}$ is $L-d$, and hence the columns of $\bfP$ form a basis of this space.
\end{proof}

\br
By Remark~\ref{ref:ORTH_CONJ_REAL}, for a real time series the basis given in Proposition~\ref{prop:ORTH_TRAJ_REPR} is a basis of $\gotL^{(L)}_{\bot}$ as well.
\er

We also note that a similar basis can be introduced for arbitrary (not only f.d.d.) time series (see \cite{HR1984})
and can be very useful in the SSA theory. In the present paper we demonstrate several applications of this basis
in the case of time series of finite difference dimension.
We will also need a (``Vandermonde''-like) basis of the trajectory space.
\bp\label{prop:IMAGE_STRUCT}
Let $F_N$ be an f.d.d. time series with the characteristic polynomial defined in \eqref{eq:CHAR_POLY}.
Let also $d < L \le N-d+1$.
Then a basis of $\gotL^{(L)}$ is given by the vectors
\bea
\ell^{0}_L(\lm_1),\ldots,\ell_L^{\nu_1-1}(\lm_1),\ldots,
\ell^{0}_L(\lm_m),\ldots,\ell_L^{\nu_m-1}(\lm_m),
\eea
where
{\rm
\bea
&\ell^{k}_L(\lm)  =  \frac{\partial^{k}}{\partial\lm^{k}} (1,\lm,\lm^2,\ldots,\lm^{L-1})^\rmT = &\\
&= (\underbrace{0,\ldots,0}_{k},k!,\ldots, \frac{(k+j)!}{j!} \lm^j,\ldots,\frac{(L-1)!}{(L-k-1)!} \lm^{L-k-1})^\rmT.&
\eea
}
\ep
\begin{proof}
Let $(G_\infty)^{\lm,k}$ be the formal continuation (as an infinite time series)
 of $\ell^k_L(\lm)$, e.g.
\bea
(G_\infty)^{\lm,0} = (1,\lm,\lm^2, \ldots).
\eea
Then by Theorem~\ref{th:LRF_FROM_P} any time series $(G_\infty)^{\lm_j,k}$, $0 \le k < \nu_j$, satisfies the LRF \eqref{eq:PME_MIN_LRF}  and all LRFs with characteristic polynomials of type $P(z)Q(z)$, where $Q(z) \not\equiv 0$. Therefore, by the second assertion of Proposition~\ref{prop:ORTH_TRAJ_REPR}, each vector $\ell^k_L(\lm_j)$ is orthogonal to 
$\gotL^{(L)}_{\bot}$. Since these $d$ vectors are linearly independent, the assertion is proved.
\end{proof}

\subsection{LRFs and continuation}\label{ssect:LRF_CONT} 
In this section we discuss the time series that can be continued within the SSA framework.
A time series $F_N$ defined in \eqref{eq:time_series_fin} \textit{admits the (forward) $L$-continuation (is $L$-continuable)}  if there exists
unique $\alpha \in \bbC$ such that $\gotL^{(L)} (F_N) = \gotL^{(L)} (f_0,\ldots,f_{N-1}, \alpha)$, see \cite[Ch. 5]{GNZH2001}
for details on continuation.
First, we show a connection between these time series and time series of finite difference dimension. 
\begin{proposition}[{\cite[Th. 5.4]{GNZH2001}}]\label{prop:CONT_SUFF_COND}
If a time series $F_N$ satisfies some LRF
\be\label{eq:LRF_CONT}
f_{n+d_0} = \suml_{k=0}^{d_0-1} a_k f_{n+k},\quad 0 \le n \le N - d_0 - 1,
\ee
with $d_0 \le \min (L-1,K)$, then it is $L$-continuable and the continuation
is achieved by the same LRF (i.e. \eqref{eq:LRF_CONT} holds for $n = N-d_0$ if we set $f_{N} = \alpha$).
\end{proposition}
\br\label{rem:CONT_FIN_INF}
By Proposition~\ref{prop:CONT_SUFF_COND} one can continue $F_N$ to an infinite
time series $F_\infty = (f_0, f_1,\ldots)$, which satisfies \eqref{eq:LRF_CONT} for all $n$ 
(i.e. $F_\infty$ is of finite difference dimension). 
This fact is the base of SSA forecasting, see \cite[Ch. 2]{GNZH2001}.
\er
The condition \eqref{eq:LRF_CONT} means that the time series $F_N$ is f.d.d.
in the sense of \cite[Ch. 2]{GNZH2001}, see also a remark after
Definition~\ref{def:FDIM_FINITE}. However, Remark~\ref{rem:CONT_FIN_INF} together with
the following assertion shows that \eqref{eq:LRF_CONT} is equivalent to the definition adopted in the present paper.

\bc\label{cor:FDIM_IS_CONT}
Any time series $F_N$ of difference dimension $d$ is $L$-continuable if  $d < L \le N-d+1$. 
\ec

Evidently, the infinite continuation (see Remark~\ref{rem:CONT_FIN_INF}) of a finite subseries $F_N$ of
an f.d.d. time series $F_\infty$ (with $d \le N/2$) coincides with the original time series $F_\infty$. 
This observation removes the ambiguity from Definition~\ref{def:FDIM_FINITE}: a finite time series $F_N$ of finite difference dimension 
cannot be a subseries of more than one infinite f.d.d. time series due to the uniqueness of continuation.

The following result, which is the converse to Proposition~\ref{prop:CONT_SUFF_COND}, can be found in \cite[Ch. 5]{HR1984}.
\bp\label{prop:CONT_NECE_COND}
If a time series $F_N$ is $L$-continuable, then there exists $d_0 \le \min (L-1,K)$ such that
$F_N$ satisfies an LRF \eqref{eq:LRF_CONT}.
\ep
\begin{proof}
The proposition is a direct consequence of \cite[Th. 5.6]{HR1984} and \cite[Prop. 5.8]{HR1984}.
\end{proof}
For convenience, we recall the well-known necessary and sufficient conditions for $F_N$ to be $L$-continuable.
\bp[{\cite[\S 5.3]{GNZH2001}}]\label{prop:CONT_EL_NESS_COND}
If $F_N$ is $L$-continuable, then $e_{L}  \notin \gotL^{(L)} (F_N)$,
where  {\rm $e_L \eqdef (0,\ldots,1)^\rmT \in \bbC^{L}$}.
\ep

\bp[{\cite[Th. 5.4]{GNZH2001}}]\label{prop:CONT_EL_COND}
Let $L \le N / 2$. If $e_{L} \notin \gotL^{(L)} (F_N)$, then $F_N$ is $L$-continuable.
\ep
We also need the notion of the backward $L$-continuation. 
\bd
A time series $F_N$ admits the \textit{backward $L$-continuation} (it is \textit{backward $L$-continuable}) if there exists
unique $\alpha \in \bbC$ such that $\gotL^{(L)} (F_N) = \gotL^{(L)} (\alpha,f_0,\ldots,f_{N-1})$.
\ed
It is clear that the backward $L$-continuation is equivalent to the forward $L$-continuation of 
the reversed time series $\what{F_N} \eqdef (f_{N-1},\ldots,f_0)$,
and the following assertion holds.
\bc
\begin{enumerate}
\item If $F_N$ is backward $L$-continuable, then $e_{1}  \notin \gotL^{(L)} (F_N)$,
where {\rm $e_1 \eqdef (1,\ldots,0)^\rmT \in \bbC^{L}$}.
\item Let $L \le N /2$. If $e_{1} \notin \gotL^{(L)} (F_N)$ then $F_N$ is backward $L$-continuable.
\end{enumerate}
\ec
We conclude this subsection with a result which is complementary to Corollary~\ref{cor:FDIM_IS_CONT} and
will be very useful in the next section.
\bp\label{prop:PME_SUFF_COND}
If $F_N$ is both forward and backward $L$-continuable, then it is a time series of finite difference dimension $d \le \min (L-1,K)$.
\ep
\begin{proof}
If $F_N$ is forward $L$-continuable then by Proposition~\ref{prop:CONT_SUFF_COND}  
there exist coefficients $a_0,\ldots,a_{d_1-1} \in \bbC$
 such that $F_N$ satisfies
\bea
f_{n+d_1} = \suml_{k=0}^{d_1-1} a_k f_{n+k}, \quad 0\le n\le N-d_1-1,
\eea
where $d_1 \le \min(L-1,K) \le N / 2$.
If $a_0 \neq 0$ then $F_N$ is an f.d.d. time series by the definition. 
If $a_0 = 0$ then we use the fact that $F_N$ is backward $L$-continuable and there exist coefficients $b_1,\ldots,b_{d_1} \in \bbC$
 such that
\bea
f_n = \suml_{k=1}^{d_2} b_k f_{n+k}, \quad 0\le n\le N-d_2-1.
\eea
Without loss of generality, assume that $d_2 \le d_1$. Then there exists a constant $\alpha \neq 0$ such that the last component $c_{d_1}$ of
\bea
(c_0,\ldots,c_{d_1})^\rmT &\eqdef& \alpha (1, -b_1,\ldots,-b_{d_2}, \overbrace{0, \ldots, 0}^{d_1-d_2})^\rmT + \\
 &+& (0,-a_1,\ldots,-a_{d_1-1},1)^\rmT
\eea
is non-zero. Then $F_N$ satisfies
\bea
f_{n+d_1} = -\suml_{k=0}^{d_1-1} \frac{c_k}{c_{d_1}} f_{n+k}, \quad 0\le n\le N-d_1-1.
\eea 
Since $c_0 = \alpha \neq 0$, the result follows from Theorem~\ref{th:POL_EXP_FROM_LRF}.
\end{proof}

\section{Separability}\label{sect:SEP}
The main result of this section is a necessary and sufficient condition of separability  
in terms of characteristic polynomials and signal roots of f.d.d. time series. In particular, it enables us 
to enumerate all possible cases of weak separability and present standard examples (see \cite[\S 6.1]{GNZH2001}) 
in a unified manner. Moreover, we develop the new theory for the \textit{one-sided separability}, 
the results on the conventional separability are its consequence.

\subsection{One-sided separability criterion}\label{ssect:SEP_LEFT}
\bd\label{def:SEP_LEFT}
Time series $F^{(1)}_N$, $F^{(2)}_N$ are called {\em weakly left-separable (right-separable)} if
$\gotL^{(L)} (F_N^{(1)}) \perp \gotL^{(L)} (F_N^{(2)})$ ($\gotL^{(K)} (F_N^{(1)}) \perp \gotL^{(K)} (F_N^{(2)})$, respectively).
\ed

For any vector $B = (b_0, \ldots,b_{L-1})^\rmT \in \bbC^{L}$
we denote by $B(z) = b_{L-1} z^{L-1} + \ldots + b_1 z + b_0$ its \textit{generating polynomial}. We 
shall also need a notation for the
polynomial with conjugate coefficients, $\ovl{B}(z) = \ovl{b_{L-1}} z^{L-1} + \ldots + \ovl{b_1} z + \ovl{b_0}$.
\bp\label{prop:POLY_COND_SEP_LEFT}
Assume that $F_N^{(1)}$ is a non-zero time series and $F_N^{(2)}$ is a time series of difference dimension $d$ with 
characteristic polynomial $P^{(2)}(z)$.
Let $L$ satisfy $d< L \le N-d+1$ and $\{U_1,\ldots,U_r\}$ be an arbitrary basis of $\gotL^{(L)} (F_N^{(1)})$.
Then $F_N^{(1)}$ and $F_N^{(2)}$ are separable if and only if 
$P^{(2)}(z)$ is a common divisor of $\ovl{U_{i}}(z)$, $1 \le i \le d$
(in other words, all the signal roots of $F^{(2)}_N$ are common roots of all $\ovl{U_{i}}(z)$, 
at least with respective multiplicities).
\ep
\begin{proof}
Note that $\gotL^{(L)} (F_N^{(1)})\perp \gotL^{(L)} (F_N^{(2)})$ if and only if
$\gotL^{(L)} (F_N^{(1)}) \subseteq \gotL^{(L)}_\bot (F_N^{(2)})$. The subspace inclusion  can be reformulated as
$\ovl{U_i} \in \gotR^{(L)} (F_N^{(2)})$ for all $1 \le i \le r$ (c.f. Definition~\ref{def:REL_SPACE}). By Proposition~\ref{prop:ORTH_TRAJ_REPR}, 
$\ovl{U_i} \in \gotR^{(L)} (F_N^{(2)})$ if and only if $\ovl{U_i} (z) = P^{(2)}(z) Q_i(z)$, 
where $P^{(2)}(z)$ is the characteristic polynomial of $F_N^{(2)}$ and $Q_i(z)$ is a nonzero polynomial. This proves the proposition.
\end{proof}
\br\label{rem:POLY_COND_SEP}
Proposition~\ref{prop:POLY_COND_SEP_LEFT} is valid even for $1< L \le N-d+1$, and thus it is valid for the essential case $L \le N/2$.
Indeed, if $L \le d$, then by Proposition~\ref{prop:POL_EXP_FIN_RANK} $\dim (\gotL^{(L)}_\bot (F_N^{(2)})) = L - L = 0$ and 
only zero time series can be left separable from $F_N^{(2)}$. At the same time, any $\ovl{U_i}(z)$ is of degree less than $d$ and hence 
it cannot be a multiple of $P^{(2)}(z)$.
\er
In what follows we assume that $L \le N/2$, unless indicated.
Proposition~\ref{prop:POLY_COND_SEP_LEFT} together with Remark~\ref{rem:POLY_COND_SEP} points out a method of finding all
f.d.d. time series $F_N^{(1)}$ that are left-separable from given $F_N^{(2)}$. 
\bp\label{prop:POLY_CRIT_SEP}
Assume that $F_N^{(1)}$ is a non-zero time series and $\{U_1,\ldots,U_r\}$ is a basis of its trajectory space $\gotL^{(L)} (F_N^{(1)})$,
$L \le N /2$.
Let $\mu_1,\ldots,\mu_l$ be all distinct nonzero common roots of $\ovl{U_1}(z),\ldots,\ovl{U_r}(z)$
and $d_1,\ldots,d_l$ be their common multiplicities, i.e. $d_k$ is the minimal multiplicity of $\mu_k$ in polynomials $\ovl{U_i}(z)$.
Then all f.d.d. time series that are left-separable from $F_N^{(1)}$ are given by
\be\label{eq:pol_exp_sign_sep}
f^{(2)}_n = \suml_{k=1}^{l} Q_{k}(n) \mu_k^n, 
\ee
where $Q_k(n)$ are (possibly zero) polynomials of degree less than $d_k$ and not all $Q_k$ are zero.
\ep
\begin{proof}
Polynomials $\ovl{U_i}(z)$ may have an additional zero root $\mu_0 = 0$ of multiplicity $d_0 \in \bbN_0$.
Then the polynomial 
\bea
R(z)= (z-\mu_1)^{d_1-1} \ldots (z-\mu_l)^{d_l-1} z^{d_0}
\eea
is the greatest common divisor of  $\ovl{U_i} (z)$. Therefore, by Remark~\ref{rem:POLY_COND_SEP},
$F_N^{(1)}$ and $F_N^{(2)}$ are left-separable if and only if the characteristic polynomial $P^{(2)}(z)$ is a divisor of $R(z)$.
By Theorem~\ref{th:POL_EXP_FROM_LRF}, the time series having this property are precisely the
 time series of form \eqref{eq:pol_exp_sign_sep}.
\end{proof}

\br
The statement of Proposition~\ref{prop:POLY_CRIT_SEP} does not depend on the choice of the basis. Indeed, if $\mu$ is a common root for the set of polynomials $\{\ovl{U_1}(z),\ldots,\ovl{U_r}(z)\}$, then it is a common root of all linear combinations of $\ovl{U_i}(z)$.
\er

Let $F_N^{(1)}$ be an f.d.d. time series  of the form \eqref{eq:pol_exp_sign}. Then a basis of  $\gotL^{(L)} (F_N^{(1)})$
is given by the vectors $\ell_L^i(\lm_k)$ from Proposition~\ref{prop:IMAGE_STRUCT}, $1 \le k \le m$, $0 \le i < \nu_k$,
and we can use Proposition~\ref{prop:POLY_CRIT_SEP} to determine all time series that are left-separable from $F_N^{(1)}$. Moreover,
we can check the left separability condition separately for each summand in the representation \eqref{eq:pol_exp_sign}.
\br\label{rem:SEP_COND_SPLIT}
Let $F_N^{(1)}$ be a time series of type \eqref{eq:pol_exp_sign}. Then an f.d.d. time series $F_N^{(2)}$
is left-separable from $F^{(1)}_N$ if and only if it is left-separable from each $P_k(n)\lm_k^n$ given in \eqref{eq:pol_exp_sign}.
\er
Indeed, each summand $P_k(n)\lm_k^n$ has the vectors $\ell_L^i(\lm_k)$, $0 \le i < \nu_k$, as a basis of its trajectory space, 
and the conditions for the signal roots of $F_N^{(2)}$ to be separable from $F_N^{(1)}$ coincide with the set
of conditions for the separability from each summand of $F_N^{(1)}$. 

\subsection{Separability of f.d.d. time series}\label{sect:SEP_FDD}
By Remark~\ref{rem:SEP_COND_SPLIT}, to study the separability of arbitrary f.d.d. time series it is sufficient to
consider the separability from a time series $P_1(n)\lm^n$, 
i.e. the time series with only one (possibly multiple) signal root. We are going to examine this case gradually
in separate examples. In the present subsection we again assume $L \le N / 2$.

\bex[Separability from a constant]\label{ex:SEP_CONST}
Let $F_N^{(1)} \equiv c \neq 0$. Then the space $\gotL^{(L)} (F_N^{(1)})$ is spanned by the single vector $(1,\ldots,1)^\rmT \in {\mathbb C}^{L}$.
Consider its generating polynomial
\be\label{eq:H_1_POLY}
W_1(z) = z^{L-1} + \ldots + 1 = \frac{z^{L}-1}{z-1}.
\ee
The roots $\mu_k$ of $W_1(z)$ are simple; they are the $L$th roots of unity, excluding 1:
\bea
\mu_k = \exp\left({\frac{2 \pi i k}{L}}\right), \quad 0 < k < L. 
\eea
In Fig.~\ref{fig:unity_roots} the roots $\mu_k$ are represented by dots.
\bfgh
\bea
 \quad \includegraphics[height=4.25cm]{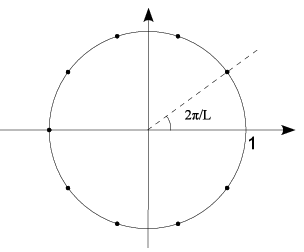}
\eea
\caption{Roots of the series left-separable from a constant.}
\label{fig:unity_roots}
\efg

By Proposition~\ref{prop:POLY_CRIT_SEP}, all left-separable from $F_N^{(1)}$ time series  have the form 
\bea
f_n^{(2)} = \suml_{k=1}^{L-1} c_k \left(\exp\left({\frac{2 \pi i k}{L}}\right)\right)^n.
\eea
Note that $F^{(2)}_N$ is an $L$-periodic series 
with 
\be\label{eq:ZERO_SUM_COND}
\suml_{n=0}^{L -1} f_n^{(2)} = 0
\ee
if and only if it has the representation as above, see \cite[\S 6.1]{GNZH2001}. In the case of real-valued time series, 
$F_N^{(2)}$ is a sum of harmonics with period $L$.
\eex

\bex[Separability from a complex exponent]\label{ex:SEP_EXP}
Let $f_n^{(1)} = \lm^n$. Then $\gotL^{(L)}(F_N^{(1)})$ is spanned by the single vector $(1,\lm, \ldots,\lm^{L-1})^\rmT$, 
and all signal roots of $F_N^{(2)}$ have to be roots of the polynomial
\bea
W_\lm(z) = \ovl{\lm}^{L-1} z^{L-1} + \ldots + \ovl{\lm} z +1.
\eea
Making the change of variables $z = x / \ovl{\lm}$, we have to solve $W_1(x) = 0$, where $W_1(x)$ is defined in \eqref{eq:H_1_POLY}.
Therefore, the roots of $W_\lm(z)$ are simple and have the form 
\bea
\mu_{k} = \frac{\exp\left({2 \pi i k}/{L}\right)}{\ovl{\lm}} = 
\frac{\lm}{|\lm|^2} \exp\left({\frac{2 \pi i k}{L}}\right),
\eea
where $0 < k < L$. Hence, the time series is of form 
\bea
f_n^{(2)} = \suml_{k=1}^{L-1} c_k \mu_k^n = \suml_{k=1}^{L-1} c_k \left(\frac{\exp\left({2 \pi i k}/{L}\right)}{\ovl{\lm}} \right)^n,
\eea
where $c_k \in \bbC$ are not simultaneously zero. 
This representation is more illustrative if we represent the roots $\mu_k$ in the polar form.
If $\lm = \rho \exp(2 \pi i\omega)$, $\rho >0$, $\omega \in [0;1)$, then 
\bea
\mu_k = \rho^{-1} \exp\left(2 \pi i\left(\frac{  k}{L}+ \omega \right)\right).
\eea
The roots $\mu_k$ are shown in Fig.~\ref{fig:sep_exp}
\bfgh
\bea
 \quad \includegraphics[height=5.4cm]{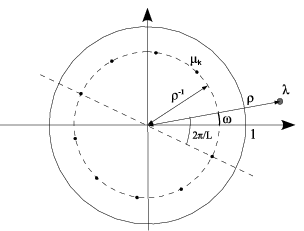}
\eea
\caption{Roots of the series left-separable from a complex exponent.}
\label{fig:sep_exp}
\efg
\eex

Let us write out the results for the case $\iim \lm = 0$, when $F_N^{(1)}$ is a real exponential time series.
If $\lm = {\rm Re\;}\lm > 0$ (i.e. $\omega = 0$), then $f^{(1)}_n = \rho^n$ is left-separable from all 
time series of type $f^{(2)}_n \eqdef \rho^{-n} f^{(3)}_n$, where $f^{(3)}_n$ has period $L$ and zero sum
(in the sense of \eqref{eq:ZERO_SUM_COND}). 
If $\lm < 0$ ($\omega = 0.5$) then $f^{(1)}_n = (-\rho)^n$ is also an exponentially modulated cosine with period $2$ (an exponentially modulated \textit{saw-tooth} series), and the set of all left-separable series can be described in a similar way.

Let us consider the important case of a sum of conjugate exponents, 
which corresponds to real cosine time series with periods other than $2$, multiplied by a complex constant. This case has some peculiarities.

\bex[Separability from a sum of conjugate exponents]\label{ex:conj_exp}
Let $f_n^{(1)} = c \lm^n + d \ovl{\lm}^n$, $\iim \lm \neq 0$. We choose $\iim \lm > 0$, without loss of generality.
Then, by Example~\ref{ex:SEP_EXP} and Remark~\ref{rem:SEP_COND_SPLIT}, any signal root $\mu$ of $F_N^{(2)}$ must satisfy
\bea
\mu = \exp\left(\frac{2 \pi i k}{L}\right) \Big/ \ovl{\lm} = \exp\left(\frac{2 \pi i l}{L}\right) \Big/ \lm,
\eea
where $0 < k,l < L$. Hence, 
\bea
\frac{\lm}{\ovl{\lm}} = \exp\left({\frac{2 \pi i m}{L}}\right),
\eea
where $m \in \mathbb{Z}$.
Therefore, $\lm$ has  to be of form
\be\label{eq:LM_CONSTRAINT}
\lm = \rho \exp\left(\frac{2\pi i m}{2L}\right),\quad \rho > 0,
\ee
for $F^{(1)}_N$ to be separable from  $F^{(2)}_N$. Since ${\rm Im}\,\lm > 0$ we can choose $m$ such that $0 < m < L$. 
Then the signal roots of $F^{(2)}_N$ are of the form
\bea
\mu_k = \rho^{-1} \exp\left(\frac{2\pi i}{2L} (2 k - m) \right), 
\eea
where $0 <  k < L$, $k \neq m$, and the time series has the representation
\bea
f_n^{(2)} = \suml_{k\in\{1,\ldots,L-1\} \setminus \{m\}} c_k \rho^{-n} \exp\left(\frac{2\pi i}{2L} (2 k - m) \right)^n.
\eea

Note that, in contrast to Example~\ref{ex:SEP_EXP}, the constraint \eqref{eq:LM_CONSTRAINT} on the signal roots of $F^{(1)}_N$ naturally arises.
In the real-valued case, if two exponentially modulated cosine series are left-separable, then these cosines should have integer periods related to $L$.
Indeed, if $f_n^{(1)} = \rho^n \cos (\varphi + 2 \pi \omega n) = c \lm^n + d \ovl{\lm}^n$ where $\varphi \in \bbR$, $\rho > 0$ and
$\omega \in (0,0.5)$,  then $\omega$ must satisfy $\omega = m / 2L$, $0 < m < L$, for $F_N^{(1)}$ to be separable from another time series.
Thus all real-valued time series $F_N^{(2)}$ that are left-separable from $F_N^{(1)}$ have the following representation
\bea
f_n^{(2)} = \!\! \suml_{k\in \{1\,\ldots,L-\delta\}\setminus\{k_0\}} c_k \rho^{-n} \cos \left(\varphi_k + 2 \pi \frac{(2k + \delta)}{2 L} n\right),
\eea
where $k_0 = \left\lfloor\frac{m}{2}\right\rfloor$,
\bea
\delta = 
\begin{cases}
0, & m \;\mbox{is even}, \\
1, & m \;\mbox{is odd},
\end{cases}
\eea
$c_k$, $\varphi_k$ are arbitrary reals and not all $c_k$ are zero.
This is in agreement with considerations in \cite[\S 6.1]{GNZH2001}, but the present exposition is more illustrative.

Note that we have two slightly different situations for even and odd $m$. In essence,
$\rho^n f^{(2)}_n$ has to be $2L$ periodic; however, for even $m$ it necessarily has the period $L$. 
This effect has been already observed (for the cosine time series) in \cite[\S 6.1]{GNZH2001},
but in the present paper, again, the nature of this effect is more evident.
Examples of locations of signal roots of $F^{(2)}_N$ for these two situations and $\rho = |\lm| = 1$ are depicted 
in  Fig.~\ref{fig:sep_conj_exp}.

\bfgh
\bea
\includegraphics[height=3.75cm]{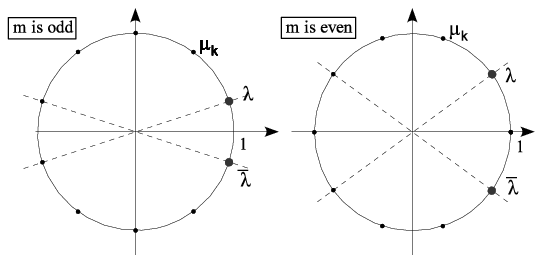}
\eea
\caption{Roots of the series (small dots) separable from a sum of conjugate exponents (large dots), $\rho = 1$.}
\label{fig:sep_conj_exp}
\efg

\eex

Finally, let us show that in the case of multiple roots no time series is separable from a given time series.

\bex[Separability from an exponentially modulated polynomial]\label{ex:poly}
Let $f_n^{(1)} = P_1(n) \lm^n$, $\lm \neq 0 $, and $P_1(n)$ be a polynomial of nonzero degree.
Then, by Proposition~\ref{prop:IMAGE_STRUCT}, 
the basis of $F_N^{(1)}$ includes the vectors 
\bea
W_\lm = \ell^0_L(\lm) = (1,\lm, \ldots,\lm^{L-1})^\rmT,\\
\ell^1_L(\lm) = (0,1, 2\lm, \ldots,(L-1)\lm^{L-2})^\rmT.
\eea
Note that in this basis we can replace $\ell^1_L(\lm)$ by 
\bea
W^{(1)}_\lm = \ell^0_L(\lm) + \lm \ell^1_L(\lm) =(1,2\lm, 3\lm^2, \ldots,L\lm^{L-1})^\rmT.
\eea
By Proposition~\ref{prop:POLY_CRIT_SEP}, all signal roots of $F_N^{(2)}$ should be at least common roots 
of the generating polynomials $W_\lm(z)$ and $W^{(1)}_\lm(z)$. Making the change of variables $z = x / \ovl{\lm}$,
 as in Example~\ref{ex:SEP_EXP}, we have to solve 
\bea
W_1(x) = W^{(1)}_1(x) = 0,
\eea
where $W_1(x)$ is defined in \eqref{eq:H_1_POLY}
 and $W^{(1)}_1(x) = L x^{L-1} + \ldots + 3 x^2 + 2x + 1$. $W_1(x)$ has the roots on the unit circle,
whereas $W^{(1)}_1(x)$ has decreasing positive coefficients, and therefore, by the Enestr\"{o}m-Kakeya theorem \cite{GR1968}, its roots
lie strictly inside the unit disc, and hence $W_1(x)$ and $W^{(1)}_1(x)$ have no common roots.
Thus, no f.d.d. time series is left-separable from $F_N^{(1)}$.
\eex

\subsection{Enumeration of separability cases}
First, let us explicitly enumerate all cases of the separability between nonzero f.d.d. time series. 
By Example~\ref{ex:poly}, only time series with the simple signal roots can be left-separable. Consider two time series
\be\label{eq:SEP_ALL}
\begin{array}{c}
f^{(1)}_n = \suml_{k=1}^r c_k \lm_k^n,\\
f^{(2)}_n = \suml_{j=1}^l d_j \mu_j^n,
\end{array}
\ee
such that $c_k,d_j \neq 0$ and $\lm_k \neq \lm_i$, $\mu_j \neq \mu_m$ for $k \neq i$, $j \neq m$. 
Then, by Remark~\ref{rem:SEP_COND_SPLIT}, these time series are left-separable if and only if each summand $c_k \lm_k^n$ 
is separable from each $d_j \mu_j^n$. Summarizing all the above and using Example~\ref{ex:SEP_EXP}, we obtain the following proposition.
\bp\label{prop:SEP_ALL}
Two f.d.d. time series  $F^{(1)}_N$ and $F^{(2)}_N$ are left-separable if and only if they
have the form \eqref{eq:SEP_ALL}, and there exist $\rho > 0$ and $\omega \in [0;1/L)$ such that 
\bea
\lm_k = \rho \exp\left(2 \pi i\left(\frac{  m_k}{L}+ \omega \right)\right), \\
\mu_j = \rho^{-1} \exp\left(2 \pi i\left(\frac{n_j}{L}+ \omega \right)\right),
\eea
where $0 \le m_k, n_k < L$ are distinct numbers.
\ep
As in Example~\ref{ex:conj_exp}, Proposition~\ref{prop:SEP_ALL} can be specialized to real-valued time series.
In this case a constraint on $L$ appears again (either $\omega = 0$ or $\omega = 1 / 2L$).

Let us examine the separability of backward and forward non-continuable  time series. Note that hereafter we do not assume $L \le N/2$.
\bp\label{prop:SEP_NCONT}
Let $F^{(1)}_N$ and $F^{(2)}_N$ be two nonzero time series, $1 < L < N$, and $e_L \in \gotL^{(L)} (F_N^{(1)})$.
These time series are left-separable if and only if both are ``border'' time series, specifically,
\be\label{eq:border}
\begin{array}{c@{}c@{}l}
F^{(1)}_N &=& (0,\ldots,0,f^{(1)}_{N-d},\ldots,f^{(1)}_{N-1}), \\
F^{(2)}_N &=& (f^{(2)}_0,\ldots,f^{(2)}_{L-d-1},0,\ldots,0),
\end{array}
\ee
where $1 < d < L$. 
\ep
\begin{proof}
Let $e_L \in \gotL^{(L)}(F_N^{(1)})$. Then the columns $X^{(L,2)}_i$ of the trajectory matrix $\bfX^{(L)} (F^{(2)}_N)$ 
satisfy $X^{(L,2)}_i \bot e_L$, $1\le i \le K$. Hence the last component of each vector $X^{(L,2)}_i$ is zero and finally $F^{(2)}_N = (f^{(2)}_0,\ldots,f^{(2)}_{L-2},\underbrace{0,\ldots,0}_{K})$. 

Let us consider the reversed time series $\what{F^{(1)}_N}$ and $\what{F^{(2)}_N}$. 
Since $F^{(2)}_N$ is nonzero, we have $e_L \in \gotL^{(L)}(\what{F^{(2)}_N})$ and, applying the first part of the proof, we obtain 
$\what{F^{(1)}_N} = (f^{(1)}_{N-1},\ldots,f^{(1)}_{K},\underbrace{0,\ldots,0}_{K})$. 
The proof is completed by noting that the sum of the lengths of the non-zero parts at the beginning of $F^{(2)}_N$ and at the end of $F^{(1)}_N$  cannot exceed $L$; the details are left to the reader.
\end{proof}

From Propositions~\ref{prop:CONT_EL_COND} and~\ref{prop:SEP_NCONT} we immediately obtain the following result.
\bc\label{cor:SEP_NCONT}
Let $F^{(1)}_N$ and $F^{(2)}_N$ be two nonzero time series, $L \le N /2$, and either $F^{(1)}_N$ is forward non-continuable or
$F^{(2)}_N$ is backward non-continuable. Then the time series are left-separable if and only if they have the form \eqref{eq:border}. 
\ec

\br
By Proposition~\ref{prop:PME_SUFF_COND}, the complete classification of all the cases of left separability 
is given by Proposition~\ref{prop:SEP_ALL} and Corollary~\ref{cor:SEP_NCONT}.
\er

Let us finish this section with the consideration of the conventional, two-sided, separability.
Two time series are called \textit{weakly (two-sided) separable}  if they are both left- and right-separable \cite[Ch. 6]{GNZH2001}.

\bp
Corollary~\ref{cor:SEP_NCONT} remains valid if we replace the left separability with the two-sided separability, and $L$ with $\min(L,K)$.
\ep
\begin{proof}
Without loss of generality we can choose $L \le N /2$. Corollary~\ref{cor:SEP_NCONT} implies that the $\boxed{\Rightarrow}$ part takes place.
The $\boxed{\Leftarrow}$ also takes place since all time series of form \eqref{eq:border} with $L = L_0$
are of the same form for all $L \ge L_0$.
\end{proof}
Therefore, we should again consider only the cases of separability of f.d.d. time series. 
Let $F^{(1)}_N$ and $F^{(2)}_N$ be time series of difference dimension $d^{(1)}$ and $d^{(2)}$, respectively. 
Then they are separable only if their trajectory matrices do not have full column or row rank, and hence, by Proposition~\ref{prop:POL_EXP_FIN_RANK} the window length $L$ satisfies the inequality
$\max(d^{(1)},d^{(2)}) < L < N - \max(d^{(1)},d^{(2)}) + 1$.
One can see that Proposition~\ref{prop:POLY_CRIT_SEP} and Remark~\ref{rem:SEP_COND_SPLIT} can be extended to handle these window lengths and hence can be applied for both $L$ and $K$. 
\bp
Let $L^*$ denote the greatest common divisor of $L$ and $K$. 
All examples from Section~\ref{sect:SEP_FDD} and Proposition~\ref{prop:SEP_ALL} are valid if we replace the left separability with 
the (two-sided) weak separability
and $L$ with $L^*$.
\ep
\begin{proof}
It is sufficient to prove the proposition only for Example~\ref{ex:SEP_CONST}, since the other assertions are based on it.
In this case the roots $\mu_k$ from Example~\ref{ex:SEP_CONST} have to be common roots of $W_1(z)$ defined in
\eqref{eq:H_1_POLY} and the polynomial
$z^{K-1} + \ldots + 1 = (z^{K}-1)/(z-1)$.
One can immediately observe that $\mu_k$ are of form 
\bea
\mu_k = \exp\left({\frac{2 \pi i k}{L^*}}\right), \quad 0 < k < L^*,
\eea
which completes the proof.
\end{proof}

\section{Extraneous roots of SSA continuation LRF}\label{sect:SSA_LRF}

In the present section we study the behaviour of the extraneous roots for the specific LRF which is used in SSA forecasting, see \cite[\S 5.2]{GNZH2001}. For an f.d.d. time series we express this LRF through the characteristic polynomial of the time series. Then we show
the correspondence between the extraneous roots and a special system of orthogonal polynomials. 
Using this correspondence, first, we demonstrate that several main properties of the extraneous roots are easily proved
and, second, we derive the asymptotic behaviour of the extraneous roots for the f.d.d. time series in the noise-free case.

\subsection{SSA LRF and its basic properties}
Let $\Lambda$ be a subspace of $\bbC^L$ such that $e_L \not\in \Lambda$.
 Let $\{U_1, \ldots, U_d\} \subseteq \bbC^L$ be the orthonormal basis of $\Lambda$ and
$U_k = \left(\begin{smallmatrix}U^\up_k \\ \pi_k \end{smallmatrix}\right)$ where $U^\up_k \in \bbC^{L-1}$ and $\pi_k \in \bbC$.
Define $\calR = (a_0,\ldots, a_{L-1})^\rmT \in \bbC^{L-1}$ as
\be\label{eq:SSA_LRF}
\calR=\frac{1}{1-\nu^2}\suml_{k=1}^d \pi_k \ovl{U^\up_k}, \quad
\ee
where \;$\nu^2 = |\pi_1|^2 + \ldots + |\pi_d|^2 = \suml_{i=1}^d |\langle U_i, e_L\rangle|^2 < 1$.
The last inequality holds since $e_L \not\in \Lambda$. The following proposition is a version of \cite[Ch. 5, Th. 5.2]{GNZH2001}
for complex-valued time series.
\bp\label{prop:SSA_LRF_FROM}
Let $F_N$ be a time series of difference dimension $d$, $1 \le d < L$.
Let $\Lambda = \gotL^{(L)}(F_N)$. Then $F_N$ satisfies an LRF
\be\label{eq:SSA_LRF_F}
f_{n+L} = \suml_{k=0}^{L-1} a_k f_{n+k},\quad 0 \le n \le N - L - 1,
\ee
with the coefficients $a_k$ given by \eqref{eq:SSA_LRF}.
\ep
The LRF \eqref{eq:SSA_LRF_F} is abbreviated as the \textit{SSA LRF}. Proposition~\ref{prop:SSA_LRF_FROM} provides the base of
the SSA continuation algorithms, see \cite[Ch. 2]{GNZH2001}. In these algorithms one finds $\Lambda$ that  
approximates $\gotL^{(L)}(F_N)$ and employs the SSA LRF obtained from $\Lambda$, which 
approximates the SSA LRF obtained from $\gotL^{(L)}(F_N)$. A detailed discussion of this approximation
can be found in \cite{G2010, N2010}.

Let $\rmR$ be the conjugate to the orthogonal complement of $\Lambda$ (c.f. Definition~\ref{def:REL_SPACE}). 
The next proposition is a version of \cite[Ch. 5, Prop. 5.5]{GNZH2001} for the complex-valued case.
\bp\label{prop:SSA_LRF_PROJ}
The vector 
{\rm
\be\label{eq:SSA_LRF_VEC}
A = (-\calR^\rmT,1)^\rmT = (-a_0,\ldots,-a_{L-1},1)^\rmT,
\ee
}
with $\calR$ given by \eqref{eq:SSA_LRF},
can be expressed as
{\rm
\be\label{eq:SSA_LRF_PROJ}
A =  c\, \Pi_{\rmR}\,e_L ,
\ee
}
where {\rm $\Pi_{\rmR}$} is the orthogonal projector on the space {\rm $\rmR$} 
and {\rm $c = (1 - \nu^2)^{-1} =  \langle\Pi_\rmR\, e_L, e_L\rangle^{-1}$ }. 
\ep
By Proposition~\ref{prop:SSA_LRF_PROJ}, 
the SSA continuation vector \eqref{eq:SSA_LRF_VEC} is equivalent to the Min-Norm prediction vector \cite{K1983, KFP1992}.
One of its well known properties, providing the name \textit{Min-Norm}, is the following; the proof can be found, 
for example, in \cite{KT1982}.
\bp\label{prop:MINNORM}
The vector \eqref{eq:SSA_LRF_VEC} yields the minimum of
\be\label{eq:SSA_LRF_PROB}
|a_0|^2 + |a_1|^2 + \ldots + |a_{L-1}|^2
\ee
among the vectors from {\rm $\rmR$}.
\ep

\subsection{SSA LRF and the characteristic polynomial}
Let us fix a time series $F_N$ of difference dimension $d$ with characteristic polynomial $P(z)$.
Let $L$ be such that $d < L \le N-d+1$ and $\Lambda =  \gotL^{(L)} (F_N)$, i.e. the noise-free case is treated. 
Then $\Pi_{\gotR} = \Pi_{\rmR}$ is the projector on the relations space $\gotR$, see Definition~\ref{def:REL_SPACE}. 
Consider the vector $B = \Pi_{\gotR} e_L$.
By Proposition~\ref{prop:SSA_LRF_PROJ}, $A = cB$. Let $A(z)$, $B(z)$ denote the generating polynomials of
$A$, $B$, see the notation in Section~\ref{ssect:SEP_LEFT}.
By Proposition~\ref{prop:ORTH_TRAJ_REPR} we obtain
\be\label{eq:H_N_DEF} 
A(z)/c = B(z) = P(z) H_n(z),
\ee
where $H_n(z) = h^{(n)}_n z^n + \ldots + h^{(n)}_1 z + h^{(n)}_0$ is a polynomial of degree  $n+1$, $n \eqdef L-d-1$. 
The $n$ extraneous roots of $A(z)$ are exactly the roots of the polynomial $H_n(z)$. 
Below we study the properties of these polynomials.

\bp {\rm
\textit{The vector} 
\be\label{eq:H_N_COEF}
H_n = (h^{(n)}_0,\ldots,h^{(n)}_n)^\rmT
\ee
\textit{is given by}
\bea
H_n = (\bfP^{*} \bfP)^{-1} e_{n+1}
\eea
\textit{where {\rm $\bfP = \bfP^{(L)}$} is defined in \eqref{eq:MULT} and {\rm $\bfP^{*}$} denotes the Hermitian conjugate of {\rm $\bfP$}.}
\textit{In other words, $H_n$ is the unique solution of}
\be\label{eq:EXT_POLY_EQ}
\bfT_n H_n = e_{n+1},
\ee
\textit{where}
\be\label{eq:DEF_T}
\bfT_n \eqdef \bfP^{*} \bfP \in \bbC^{(n+1)\timess (n+1)}.
\ee}
\ep
\begin{proof}
By Propositions~\ref{prop:SSA_LRF_PROJ} and~\ref{prop:ORTH_TRAJ_REPR}, $B = \bfP^{(L)} H_n$ and 
$H_n$ minimizes the Euclidean norm $\|\bfP^{(L)} V - e_L\|_2$ among all vectors $V \in \bbC^{L-d}$. 
Therefore, $H_n$ can be expressed as the least squares solution of $\bfP^{(L)} V \approx e_L$:
\be\label{eq:EXT_POLY_1}
H_n = (\bfP^{*} \bfP)^{-1} \bfP^{*} e_{L} = (\bfP^{*} \bfP)^{-1} e_{n+1},
\ee 
where the last equality holds since $p_d = 1$.
\end{proof}

\br
In fact, the projector $\Pi_{\gotR}$ itself can be explicitly expressed through the characteristic polynomial as
$\Pi_{\gotR} = \bfP (\bfP^{*} \bfP)^{-1} \bfP^{*}$.
\er

By construction, the matrix $\bfT_n$ defined in \eqref{eq:DEF_T} is a Toeplitz Hermitian matrix
\bea
\bfT_n =  (t_{i-j})_{i,j=0}^{n, n}
\eea
with 
\be\label{eq:COV_MATR_COEF}
t_k = 
\begin{cases}
\suml_{j=0}^{d-k} \ovl{p_j} p_{k+j}, & 0 \le k \le d, \\
0, & k > d, \\
\ovl{t_{-k}}, & k < 0.  \\
\end{cases}
\ee
Note that the coefficients $t_k$ do not depend on $n$. 
Also, any matrix $\bfT_n$ has no more than $2 d + 1$ non-zero diagonals 
\bea
\bfT_n= 
\begin{pmatrix}
t_0    & \dots  & t_{-d} &        &  \\
\vdots & \ddots & \ddots & \ddots &  \\
t_d    & \ddots & t_0    & \ddots & t_{-d} \\
       & \ddots & \ddots & \ddots & \ddots &  \\
       &       & t_d    & \ddots & t_0 \\
\end{pmatrix}.
\eea

By the way, $t_k$ coincide with the values of the covariance function for
 the moving average process with coefficients $p_k$, see \cite[Ch. 5]{A1971}.
If we rewrite the equation \eqref{eq:EXT_POLY_EQ} as
\be\label{eq:YW_EQ}
\bfT G_n = e_1,
\ee
where $G = (\ovl{h^{(n)}_n},\ldots, \ovl{h^{(n)}_0})^\rmT$, then we obtain the Yule-Walker equation for this process.

\subsection{Orthogonal polynomials. Basic properties}
In this subsection we apply the powerful theory of orthogonal polynomials on the unit circle and obtain short proofs
for the basic properties of extraneous roots. Let us first rewrite the equation
\eqref{eq:EXT_POLY_EQ} in terms of orthogonal polynomials. Let  $t(z) = \suml_{k=-d}^d t_k z^k$, 
where $t_k$ are given by \eqref{eq:COV_MATR_COEF}. It is easy to see that $t(z) = P(z) \ovl{P}(1/z)$ and the relation
\be\label{eq:w_repres}
t(z) = P(z) \ovl{P}(\ovl{z}) = |P(z)|^2 \ge 0
\ee
holds for all $z\in \Ccircle_1$, where $\Ccircle_r = \{z \in \bbC: |z|=r\}$ denotes the circle of radius $r$. 
 
For a non-negative function $w(z)\in \bfL_1 (\Ccircle_1)$ (i.e. a Lebesgue integrable over the
 contour $\Ccircle_1$ function) one can 
define the inner product in the space of complex polynomials
\be\label{eq:def_orth2}
\left\langle p(z), q(z) \right\rangle_w \eqdef
\frac{1}{2 \pi} \intl_{-\pi}^{\pi} p(z) \ovl{q(z)} w(z) d\theta,
\ee
where $z = e^{i\theta}$. The function $w(z)$ is called the \textit{weight}. In particular, 
one can define the inner product $\langle\cdot,\cdot\rangle_t$ for the weight $t(z)$.
\bp\label{prop:H_N_ORT}
Polynomials $H_n(z)$, $n \ge 0$, defined by \eqref{eq:H_N_DEF} form an orthogonal system with respect to $\langle\cdot,\cdot\rangle_t$.
\ep
\begin{proof}
It is evident that $t_{k-l} = \left\langle 1, z^{k-l} \right\rangle_t = \left\langle z^l, z^k \right\rangle_t$ for all $k,l \in \bbZ$. 
Therefore, we can rewrite the equation \eqref{eq:EXT_POLY_EQ} as 
\bea
\left\langle H_n(z), z^k \right\rangle_t = 
\suml_{l=0}^n h^{(n)}_l t_{k-l}  =
\begin{cases}
0, & 0 \le k < n, \\
1, & k = n.
\end{cases}
\eea
Hence $\left\langle H_n(z), H_m(z) \right\rangle_t = 0$ for $m \neq n$ and
\bea
 \| H_n(z)\|^2_t \eqdef \left\langle H_n(z), H_n(z) \right\rangle_t = h^{(n)}_n \neq 0
\eea
for all $n \ge 0$, where $h^{(n)}_n$ is given in \eqref{eq:H_N_COEF}, and the assertion is proved.
\end{proof}

\br\label{rem:ORTH_UNIQUE}
Let $\Phi_n(z)$, $n\ge 0$, be a system of polynomials of degree $n$, which are orthogonal with respect to some weight $w(z)$ 
(i.e. $\left\langle \Phi_n(z), \Phi_m(z) \right\rangle_w = 0$ for $m \neq n$, $\|\Phi_n(z)\|^2_w \neq 0$ for all $n$). Then $\Phi_n$ 
are defined uniquely up to constant factors. This follows from the properties of the standard orthogonalization process applied 
to the sequence $\{1,x,x^2, \ldots\}$, see \cite[Ch. IX, \S 6]{G1998}.
\er
Thus, Proposition~\ref{prop:H_N_ORT} is a characterization of $H_n(z)$. 
\br\label{rem:ORTH_MUL_CONST}
The system $\Phi_n$ from Remark~\ref{rem:ORTH_UNIQUE} is also orthogonal with respect to the weight $\alpha w(z)$ for all $\alpha >0$.
\er
Now let us recall the well-known property of orthogonal polynomials, which has a consequence for the extraneous roots.
We provide  the proof just to show its simplicity, it also can be found in {\cite[Ch. 1]{S2005}}.
\bt\label{th:roots_prop1}
Let $\Phi_n(z)$ be an orthogonal polynomial of degree $n$ with respect to $\langle\cdot,\cdot\rangle_w$.
Then $|z_0| < 1$ if $\Phi_n(z_0) = 0$, i.e.
the roots of the polynomial $\Phi_n$ are located inside the unit circle.
\et
\begin{proof}
If  $\Phi_n$  has a root $z_0$ then $\Phi_n(z) = Q(z) (z - z_0)$, where $\deg Q = n-1$, and hence $\langle \Phi_n, Q\rangle_w = 0$. Then
\bea
&\|Q(z) \|^2_w = \langle z Q(z), z Q(z)\rangle_w  = \|z Q(z) \|^2_w = &\\
 & =\|Q(z) z_0 + \Phi_n(z) \|^2_w =  |z_0|^2 \|Q(z)\|^2_w + \|\Phi_n(z) \|^2_w, &
\eea
where the last equality follows from the orthogonality of polynomials. Rewriting we have $(1- |z_0|^2)\|Q(z)\|^2_w = \|\Phi_n(z) \|^2_w > 0$, 
which completes the proof.
\end{proof}
In particular, we obtain the following result  for the SSA LRF (Min-Norm prediction), which is well known  \cite{KT1982,K1983,BB1998}, but the proof is often too complicated.
\bc
All extraneous roots of SSA LRF lie inside the unit circle.
\ec

Next, we provide another clear proof for the well-known result \cite{BB1998} on the correspondence
 between the extraneous roots for the forward and backward SSA LRFs.
By the backward SSA LRF we mean the SSA LRF for the reversed time series $\what{F_N}$; the forward LRF means the standard SSA LRF.
It is easy to observe (for example, using the basis in Proposition~\ref{prop:ORTH_TRAJ_REPR}) that  $\what{F_N}$ is
an f.d.d. time series with the characteristic polynomial $p_0^{-1}\what{P}(z)$,
where $\what{P}(z)$ denotes the reversed polynomial $p_d + p_{d-1} z + \ldots + p_0 z^d$.
This polynomial has roots $\lm_k^{-1}$ with multiplicities $\nu_k$ (c.f. \eqref{eq:CHAR_POLY}).
For convenience, for any polynomial $B(z) = b_r z^r + \ldots + b_1 z + b_0$, $b_r \neq 0$, we introduce the notation
\be\label{eq:POLY_CONJ_DEF}
B^*(z) \eqdef \what{\ovl{B}}(z) = \ovl{b_0} z^r + \ldots + \ovl{b_{r-1}} z + \ovl{b_r}.
\ee
Then
\be\label{eq:FORW_BACK_CHAR_POLY}
|P(z)|^2 = |P^*(z)|^2 = |\what{P}(\ovl{z})|^2, \quad |z| = 1.
\ee
This fact enables us to easily show the following.
\bp\label{prop:back_forw_lrf}
The extraneous roots of the backward and forward SSA LRFs  are conjugate.
\ep
\begin{proof}
Let $u(z) = |p_0|^{-2} |\what{P}(z)|^2$ denote the weight for the reversed time series. Then
for any $Q(z)$ and $S(z)$ we obtain
\bea
\langle Q(z), S(z)\rangle_u = |p_0|^{-2} \langle \ovl{P}(z), \ovl{Q}(z)\rangle_t,
\eea
which follows from  \eqref{eq:def_orth2}. Therefore, each polynomial from a set of polynomials orthogonal with respect to $u(z)$  
has the roots which are conjugate to the roots of the corresponding $H_n(z)$, and the assertion is proved.
\end{proof}
Note that it is common to define the backward SSA LRF with conjugation of the coefficients \cite{K1983}. In this case, the assertion of Proposition~\ref{prop:back_forw_lrf} changes; specifically, the extraneous roots of the backward and forward SSA LRFs coincide (as stated in Introduction).

\subsection{Asymptotic properties}\label{ssect:asymp}
In this section we present a review of recent results on the asymptotic distribution of the roots of orthogonal 
polynomials which can be used to study the extraneous roots for the SSA LRF. We mainly follow \cite{MMS2006a} and \cite{MMS2006s}
and present all the facts in a unified and simplified manner, specializing the results for the weight $t(z)$ defined in \eqref{eq:w_repres}. 

For convenience, in addition to $\Ccircle_r$ we define
\bea
&&\Cdisk_r (a) \eqdef \{z \in \bbC : |z-a| < r\}, \\
&&\Cdisk_r \eqdef \Cdisk_r(0), \\
&&\Cdisk^c_r \eqdef  \{z \in \bbC : |z| \le r\},
\eea
for $r > 0$ and $a \in \bbC$. 

Let us consider the characteristic polynomial $P(z)$ defined in \eqref{eq:CHAR_POLY} and the orthogonal polynomials
$H_n(z)$ defined in \eqref{eq:H_N_DEF}.
Using a transformation similar to  \eqref{eq:FORW_BACK_CHAR_POLY} 
we can ``transfer'' all the roots inside the closed unit disk $\Cdisk^c_1$. Define
\be\label{eq:PRE_NORM}
C(z) = \prod\limits_{k: |\lm_k| \le 1} (z-\lm_k)^{\nu_k} 
\prod\limits_{l: |\lm_l| > 1} (z-\ovl{\lm_l^{-1}})^{\nu_l}.
\ee
Then 
\be\label{eq:reflect_root}
 |P(z)|^2  =  c |C(z)|^2,
\ee
where $c$ is some positive constant. Hence, by Remark~\ref{rem:ORTH_MUL_CONST}, the weight $t(z) = |P(z)|^2$ and the weight $u(z) = |C(z)|^2$ generate the same system of orthogonal polynomials $H_n$, and the roots of $C(z)$ are inside the closed unit disk.

Note that in the representation \eqref{eq:PRE_NORM} the pairs of roots of $P(z)$ which are related by $\lm_k = \ovl{\lm_l^{-1}}$
are glued together. The \textit{normalized representation} of $C(z)$, defined in \eqref{eq:NORM_REPR}, takes this into account. We set
\be\label{eq:NORM_REPR}
C(z) = (z-a_1)^{m_1}\ldots(z-a_s)^{m_s}
\ee
where all $a_k$, $1 \le k \le s$, are distinct. 

Let us also introduce several definitions. The radius
$\rho = \max_{1 \le k \le s} |a_k| \le 1$ is called the {\em critical radius} for the polynomial $P(z)$. The circle
$\Ccircle_\rho$ is called the \textit{critical circle}. 
Moreover, let the roots of $C(z)$ be ordered such that the first $u$ roots are on the critical circle and the other roots are inside it,
i.e. $|a_1| =  \ldots = |a_u| = \rho$  and $|a_k| < \rho$ for $k > u$. We will call $a_1, \ldots, a_u$ the \textit{leading roots} of $C(z)$.
Let also the first $\ell$ roots $a_1,\ldots,a_\ell$ be of greatest multiplicity $M$ among $a_1,\ldots, a_u$,
i.e. $m_s = M$ for all $1 \le s \le \ell$ and $m_j < M$ for $\ell <j \le u$. 
\bp[{\cite[Prop. 1, Th. 3]{MMS2006a}}]\label{prop:zeros_loc_analytic}
Let $\rho < 1$. 
\begin{enumerate}
\item For any $\varepsilon > 0$ there exists $N_1(\varepsilon)$ such that 
all zeros of $H_n(z)$ are inside $\Cdisk_{\rho + \epsilon}$ for all $n \ge N_1$.
\item For any $\varepsilon > 0$ there exists $N_2(\varepsilon)$ such that 
the closed disk $\Cdisk^c_{\rho - \varepsilon}$ contains at most $\ell-1$ roots of $H_n(z)$
for all $n \ge N_2$.
\end{enumerate}
\ep
\bp[{\cite[Cor. 1]{MMS2006s}}]\label{prop:zeros_loc_singular}
If $\rho = 1$ then for any $\varepsilon > 0$ there exists $N_2(\varepsilon)$ such that for all $n \ge N_2$
the closed disk $\Cdisk^c_{1-\varepsilon}$ contains at most $u-1$ roots of $H_n(z)$.
\ep

These two propositions imply that the majority of roots tend uniformly to the critical circle $\Ccircle_\rho$ 
(they are called \textit{general roots}), and a bounded number of roots stay strictly inside $\Cdisk_\rho$.
(called \textit{spurious} roots). The spurious roots chaotically float inside $\Cdisk_{\rho}$ (when changing $n$), 
but, roughly speaking, they are asymptotically close to zeros of the functions $G_n$ defined by
\bea
G_n(z) = 
\begin{cases}
\displaystyle \suml_{k=1}^{\ell} \frac{a_k^{n-M+d+1} C^*(a_k)}{(z-a_k) C^{(M)}(a_k)}, & \rho < 1, \\
\displaystyle \suml_{k=1}^{u} \frac{a_k^{n+d+1} (-1)^{\nu_k} \nu_k (C^*)^{(\nu_k)}(a_k)}{(z-a_k) C^{(\nu_k)}(a_k)}, & \rho = 1, \\
\end{cases}
\eea
where $f^{(m)}$ stands for the $m$-th derivative of a function $f$ and $C^*(z)$ is defined as in
\eqref{eq:POLY_CONJ_DEF}. See \cite{S1979,MMS2006a,MMS2006s} for exact formulations; see also
an example with spurious roots at the end of this section.

Let us look at the behaviour of general extraneous roots. We present only an informal summary
in order not to overload the exposition with technicalities. Precise and mathematically strict results
can be found in \cite[Th. 4]{MMS2006a} and \cite[Th. 5]{MMS2006s} (for $\rho < 1$ and $\rho = 1$, respectively).
\begin{itemize}
\item The asymptotics for the absolute values of the general roots are
\bea
|z_i^{(n)}| = 
\begin{cases}
\displaystyle \rho \left(1 + M\frac{\log(n)}{n} + O\left(\frac{1}{n}\right)\right),& \; \rho < 1, \\
\displaystyle  1 + \frac{\log(n)}{n} + O\left(\frac{1}{n}\right),& \; \rho = 1.
\end{cases} 
\eea

\item 
For a function $f$, define by $Z_\varepsilon(f) = \bigcup_{a: f(a) = 0} \Cdisk_\varepsilon(a)$ the
$\varepsilon$-vicinity of its zero set. Let us denote
\bea
\calB_{C,\varepsilon} = 
\begin{cases}
\bigcup_{k=1}^{l} \Cdisk_\varepsilon(a_k), & \rho < 1, \\
\bigcup_{k=1}^{u} \Cdisk_\varepsilon(a_k), & \rho = 1. \\
\end{cases}
\eea
Then for large $n$ and small $\delta$ there are no spurious roots in the area 
\bea
\calA_{\varepsilon,\delta,n} = \{z:  ||z| - \rho| < \delta\} \setminus (\calB_{C,\varepsilon} \cup Z_\varepsilon(G_n)).
\eea

\item
Let $\calD$ be a connected sub-area of $\calA_{\varepsilon,\delta,n}$ and $z^{(n)}_1, \ldots , z^{(n)}_r$ denote 
the roots of $H^{n}(z)$ in $\calD$, ordered by the magnitude of their arguments. 
Then for $i,j$ such that $1 \le i, (i+j) \le r$ we have
\bea
\arg(z^{(n)}_{i+j}) - \arg(z^{(n)}_{i}) = \frac{2\pi j}{n} + O \left(\frac{1}{n^2}\right).
\eea 
\end{itemize}
These  results can also be found in \cite{S1979} in a weaker form. Let us make some comments. 
\begin{itemize}
\item The general roots show an asymptotic angular equidistribution. The order of their convergence to the critical circle
 differs for $\rho = 1$ and $\rho < 1$. In the latter case the convergence rate depends on
the maximum multiplicity $M$ of the leading roots of $C(z)$. 
\item For $\rho < 1$ only the roots $a_1,\ldots,a_\ell$ 
(i.e. the leading roots  with maximum multiplicity $M$)
 significantly affect the asymptotic behaviour of extraneous roots: they determine the number of spurious roots,
and their vicinities are excluded from $\calA_{\varepsilon,\delta,n}$; however for $\rho = 1$ all leading roots of $C(z)$ 
affect the asymptotic behaviour.
\item In the real-valued case each signal root is usually accompanied with its conjugate (when the signal is a sum of modulated cosines).
Therefore, there is likely to be more than one leading root of $C(z)$, and the spurious roots are likely to appear.
\end{itemize}
 
\bex[General and spurious roots]
Consider a sum of two exponentially modulated (with the same exponent) cosine time series. 
Since each cosine corresponds to two conjugate roots, we have $\ell = u = 4$ and $M = 1$.
The roots of the SSA LRF are plotted in Fig.~\ref{fig:extsam} (recall that $L = n+d+1$).

\bfgh
\bea
\begin{array}{@{}c@{}c@{}}
\includegraphics[height=4.5cm]{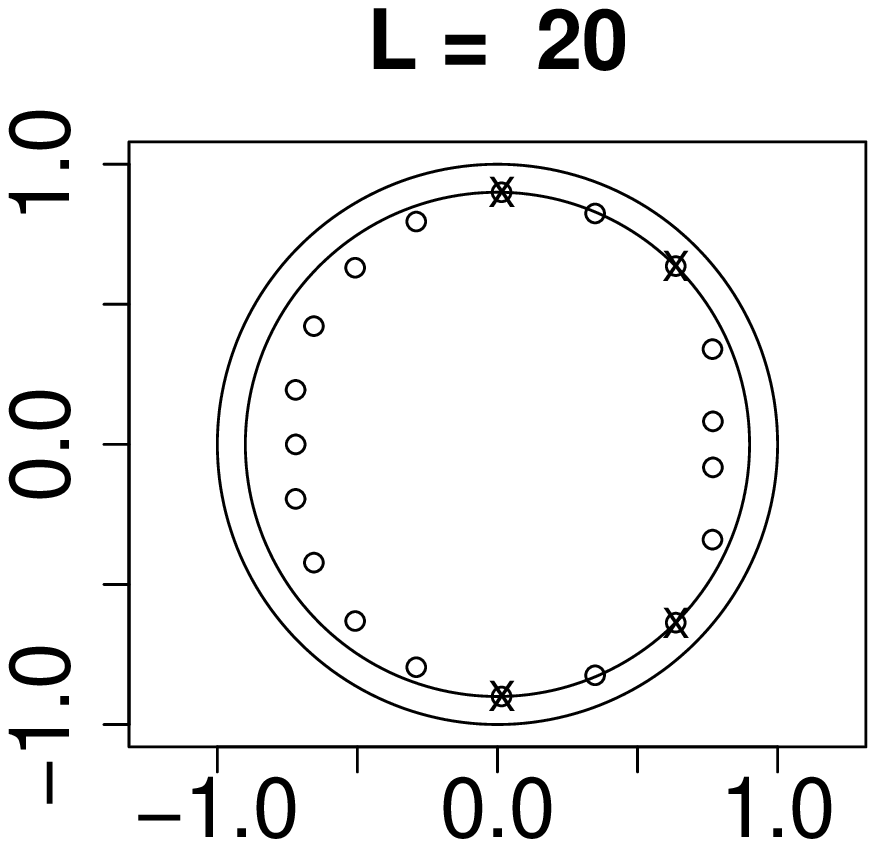}& \includegraphics[height=4.5cm]{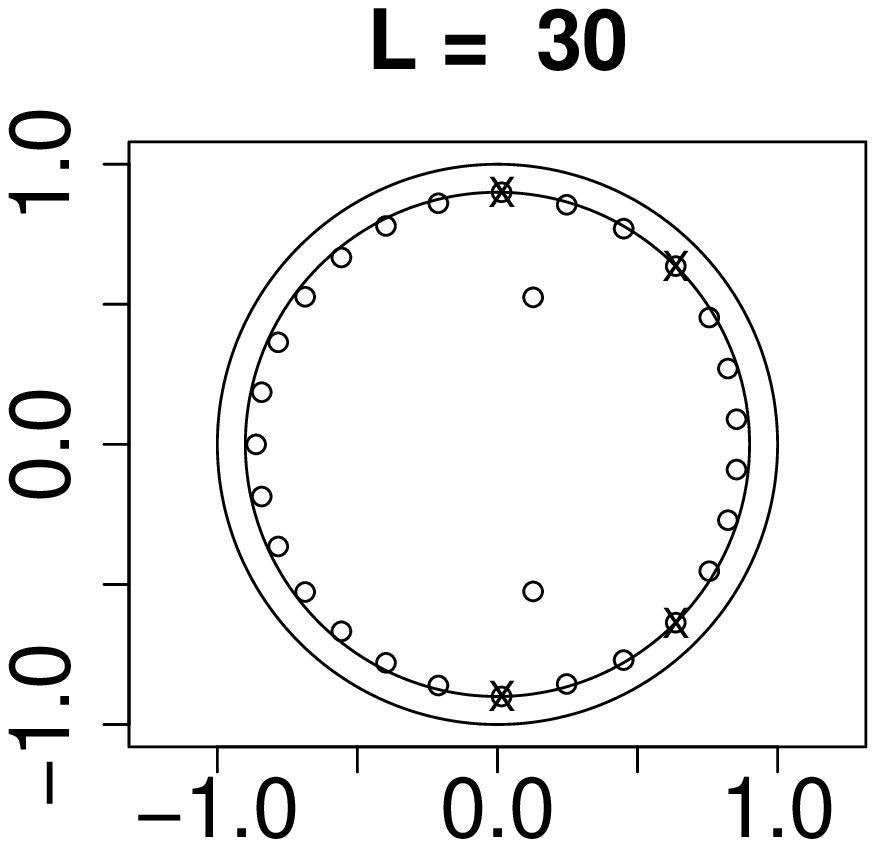}\\
\includegraphics[height=4.5cm]{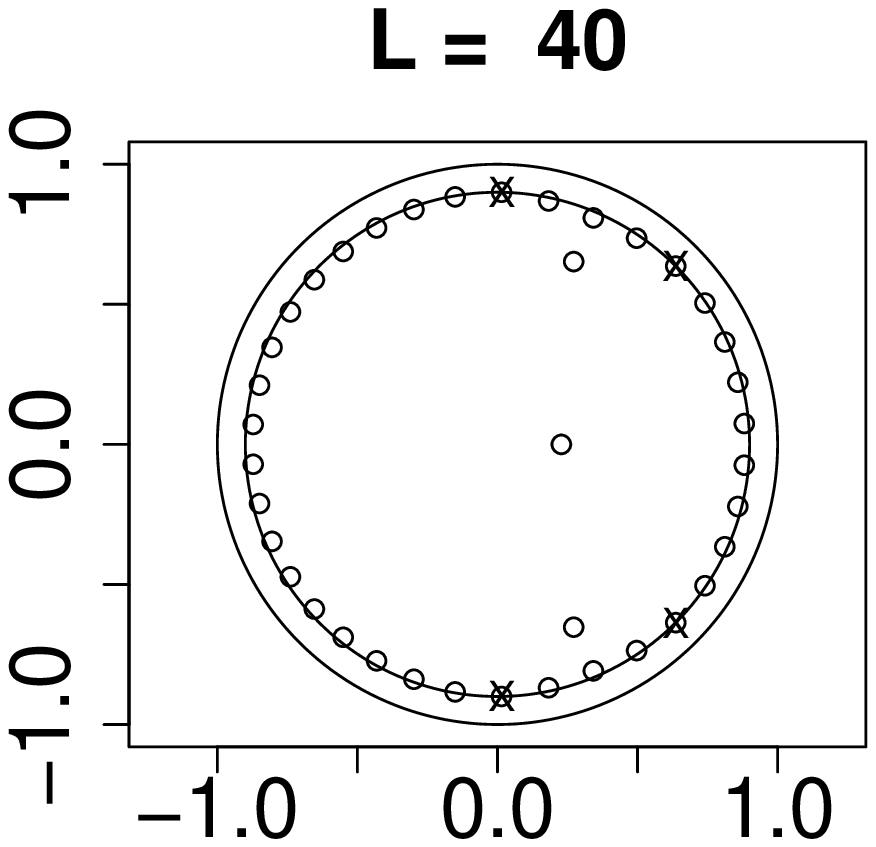}& \includegraphics[height=4.5cm]{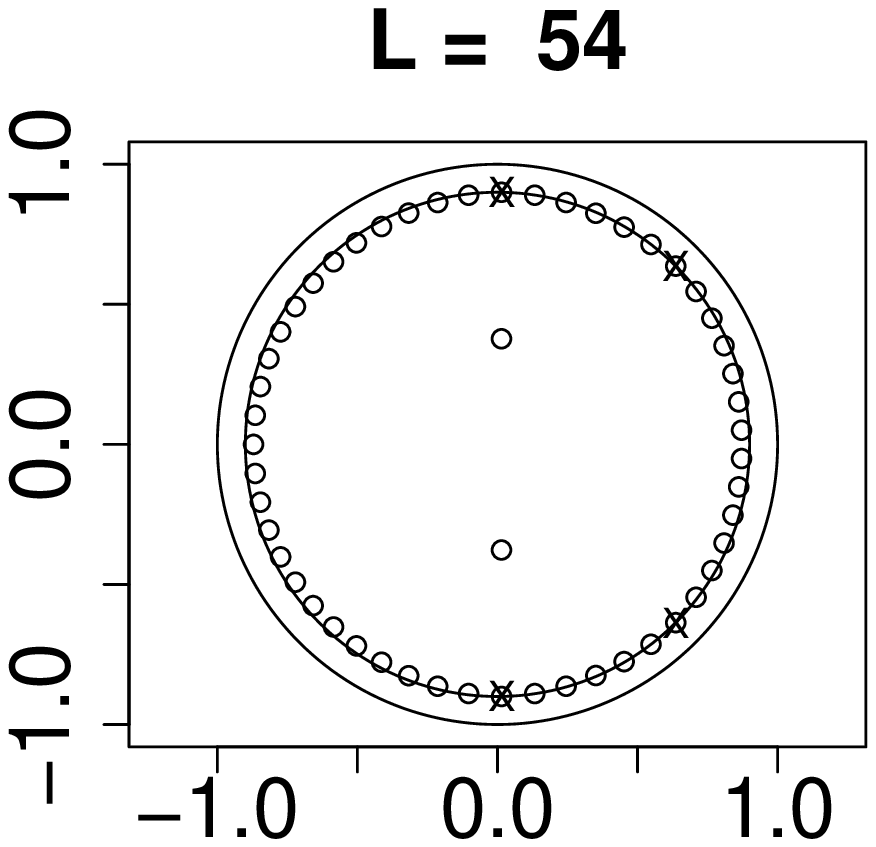}\\
\end{array}
\eea
\caption{Signal (x) and extraneous (o) roots, 
$f_n = 0.9^n\left(\cos\left( \frac{2\pi}{8} n \right) + \cos\left( 2\pi \sin(0.25) n \right)\right)$}
\label{fig:extsam}
\efg
As one can see in Fig.~\ref{fig:extsam}, the spurious roots float inside the critical circle; for some $L$ the spurious roots
disappear. 
The coefficient $\sin(0.25)$ was chosen to ensure the rational independence of the leading roots of $C(z)$. 
If the majority of leading roots is rationally dependent (i.e. some roots are expressed through linear combinations of others with rational
coefficients), then the spurious roots float more regularly, see \cite{S1979,MMS2006a,MMS2006s} for details.
\eex
 
\subsection{Several applications and remarks}
First, let us show a connection between separability and the behaviour of extraneous roots.
We demonstrate that the approximate \cite[Ch. 6]{GNZH2001} (left) separability  of a linear time series from a periodic time series 
can be informally justified by the condition of separability developed in Section~\ref{sect:SEP}. 
Let $W^{(L,0)} (x)\eqdef W_1(x) $ and $W^{(L,1)} (z)\eqdef W^{(1)}_1(z)$ be the polynomials from Example~\ref{ex:poly}. 
The roots of $W^{(L,0)}$ are always on the unit circle and uniformly distributed with equal angles
between adjacent roots on it (without number $1$). One can show that $W^{(L,1)}(x)$ for $L = 1, 2, \ldots$ form an orthogonal 
system with respect to the weight $P(z) = |z-1|^2$, and by results in Section~\ref{ssect:asymp} the roots of $W^{(L,1)}(x)$ are distributed
 asymptotically in the same way as the roots of $W^{(L,0)}(x)$.
 In Fig.~\ref{fig:linear} one can see the roots of $W^{(L,0)}(x)$ and $W^{(L,1)}(x)$ for large $L$.
\bfgh
\bea
\includegraphics[height=5cm]{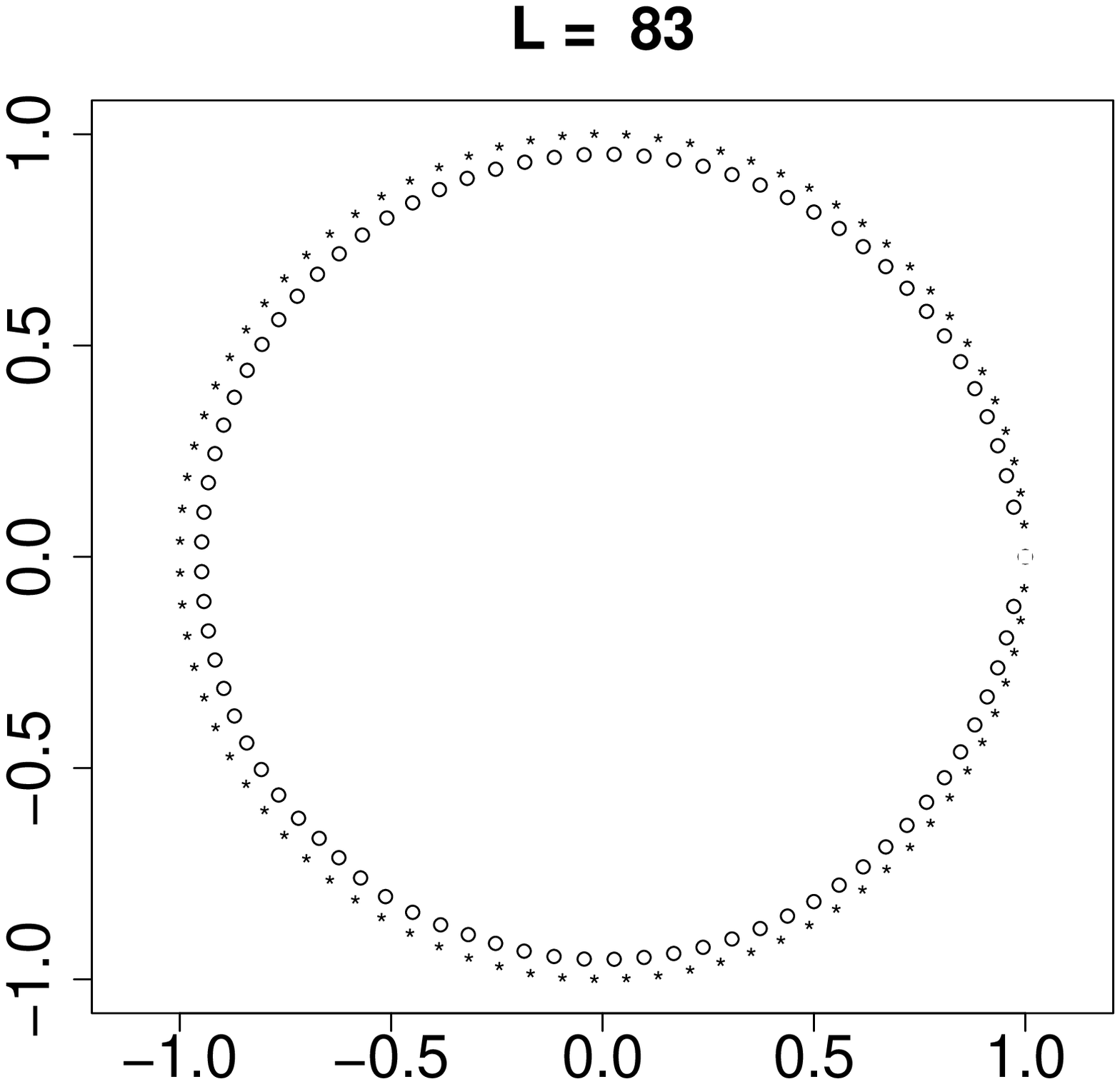}
\eea
\caption{The roots of $W^{(L,0)}$ and $W^{(L,1)}$.}
\label{fig:linear}
\efg

We might say that this validates the asymptotic separability of an arbitrary periodic series from a linear function. The same can be shown
for any polynomial time series. Note that this observation does not pretend to be a rigorous assertion. A recent theoretical study of asymptotic and approximate separability can be found in \cite{N2010}.

Let us now consider the problem of determining the signal roots from the approximate SSA LRF (the Min-Norm estimation of the main roots).
Theorem~\ref{th:roots_prop1} is the base of the \textit{root-Min-Norm} \cite{KT1982,KFP1992, SM1997} method for estimation of exponents. Under the assumption that all signal roots have modulus not less than $1$, one selects the greatest by absolute value $d$ roots from the estimated LRF to be the signal roots. 

If all signal roots have the absolute values greater than $1$, we conclude from the asymptotic distribution of the roots that the root-Min-Norm approach is applicable. In the presence of moderate noise, the main and extraneous roots of the estimated LRF are close to those of the noiseless 
LRF, see, for example, \cite{KT1982}.
In Fig.~\ref{fig:noised} a noisy time series with its SSA LRF roots are depicted.
\bfgh
\bea
\begin{array}{@{}c@{}c@{}}\includegraphics[height=4.5cm]{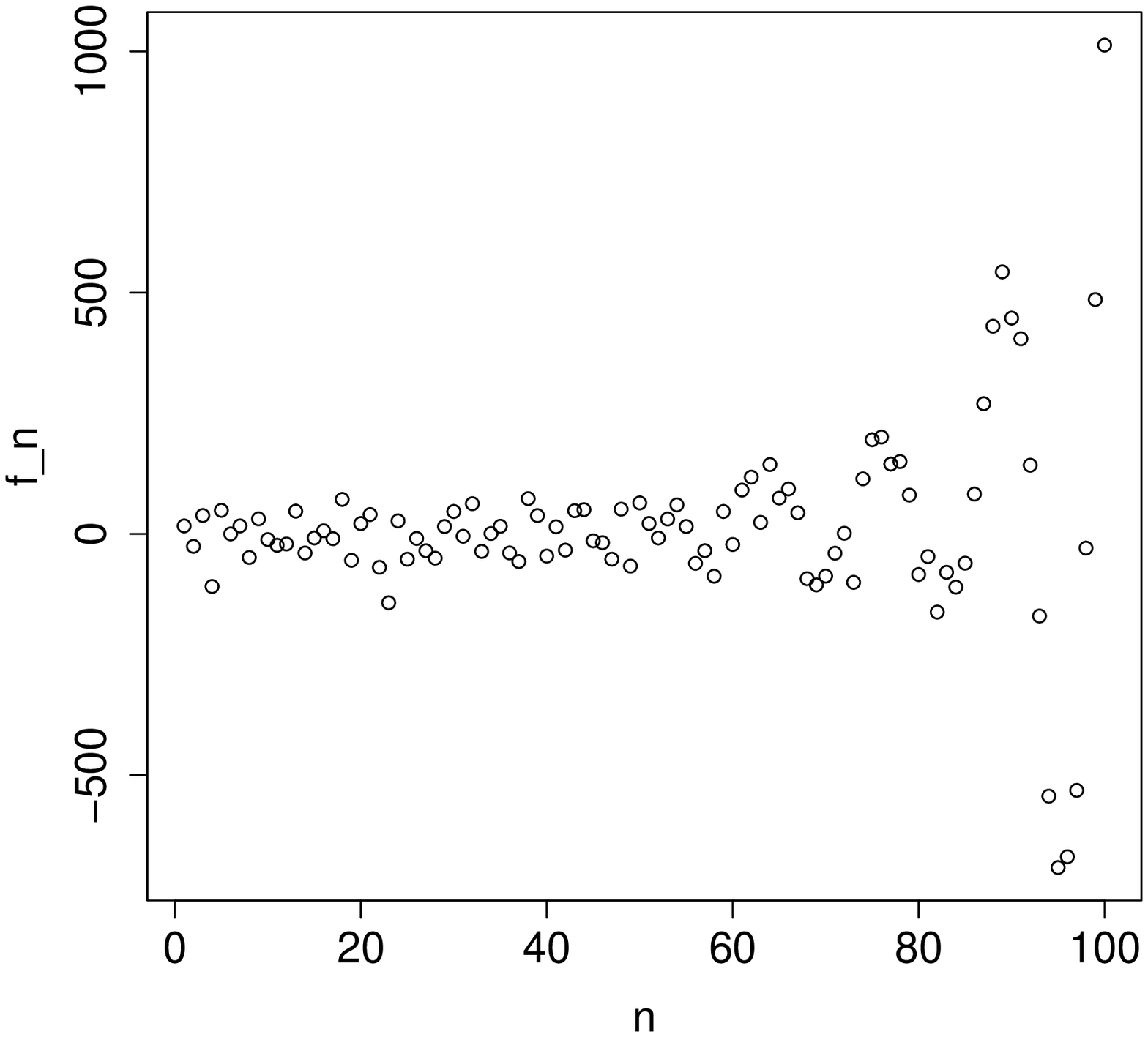}&\includegraphics[height=4.5cm]{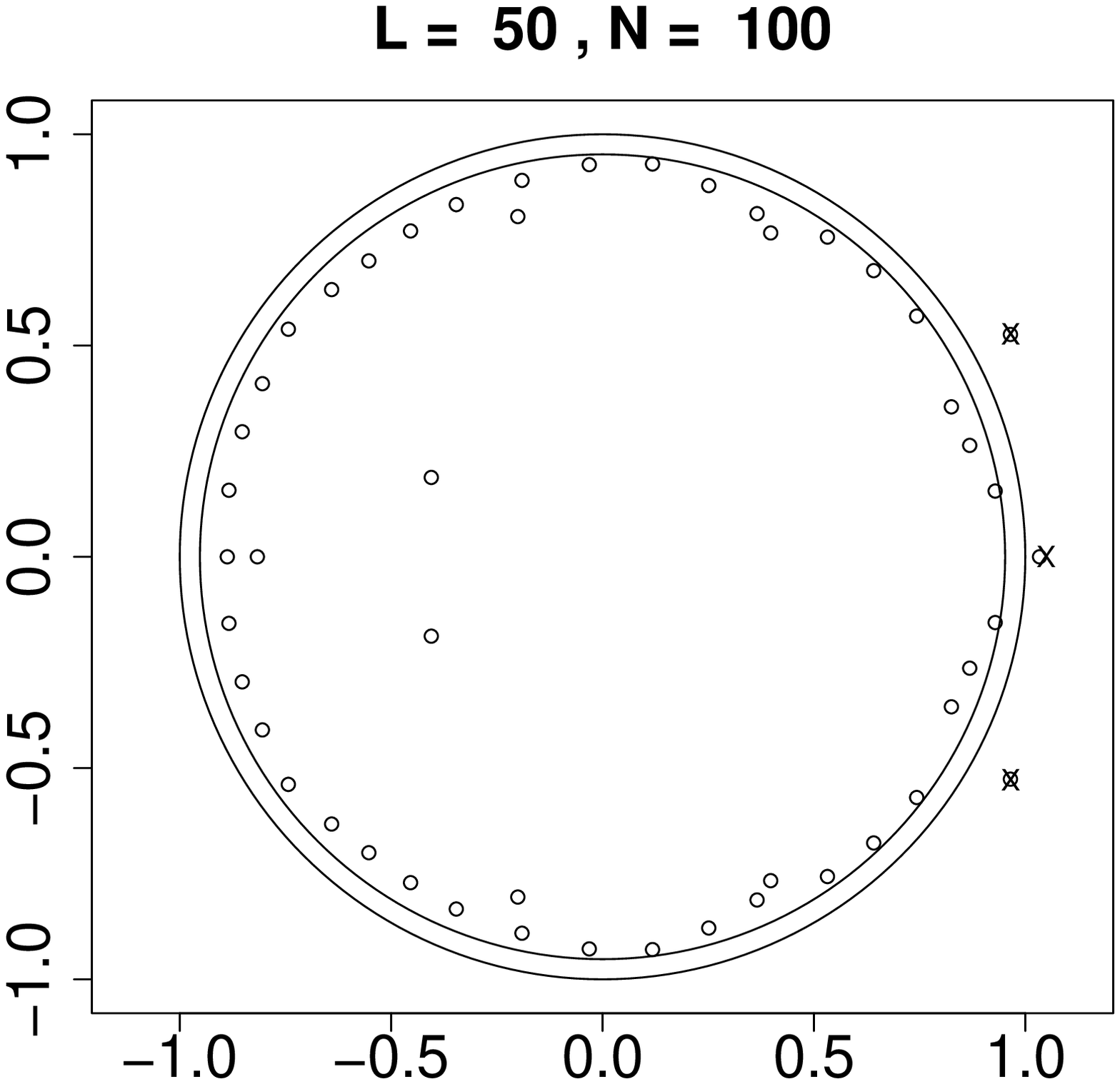}\end{array}
\eea
\caption{
Left: the time series,
Right: the signal roots (x) and the extraneous roots of the SSA LRF (o), 
 $f_n = 1.05^n + \frac{1}{10} 1.1^n \cos(0.5 n) + \varepsilon_n$, 
$\varepsilon_n$ are i.i.d. $\calN(0,50^2)$,
}
\label{fig:noised}
\efg
If some signal roots are on the unit circle, then this approach behaves worse. 

If all signal roots have modulus less than $1$, then the naive root-Min-Norm approach is not applicable. 
Even if the signal roots are only on  the critical circle, the extraneous roots may have larger modulus, as shown in Fig.~\ref{fig:mult}.
In this case one can use the backward SSA LRF to get the main and extraneous roots separated, see \cite{KT1982}.

Assume now that within signal roots there are roots both outside and inside the unit circle. In this case one
can consider the backward and forward SSA LRF simultaneously and use the relations between the extraneous and signal roots of both LRFs.

Finally, we discuss the notion of multiplicity. Consider the case of a single real exponent time series, modulated by a 
quadratic polynomial. There is one signal root of multiplicity $3$.
In Fig.~\ref{fig:mult} the roots of SSA LRF for the noise-free case and the noise case are depicted.
\bfgh
\bea
\begin{array}{@{}c@{}c@{}}\includegraphics[height=4.6cm]{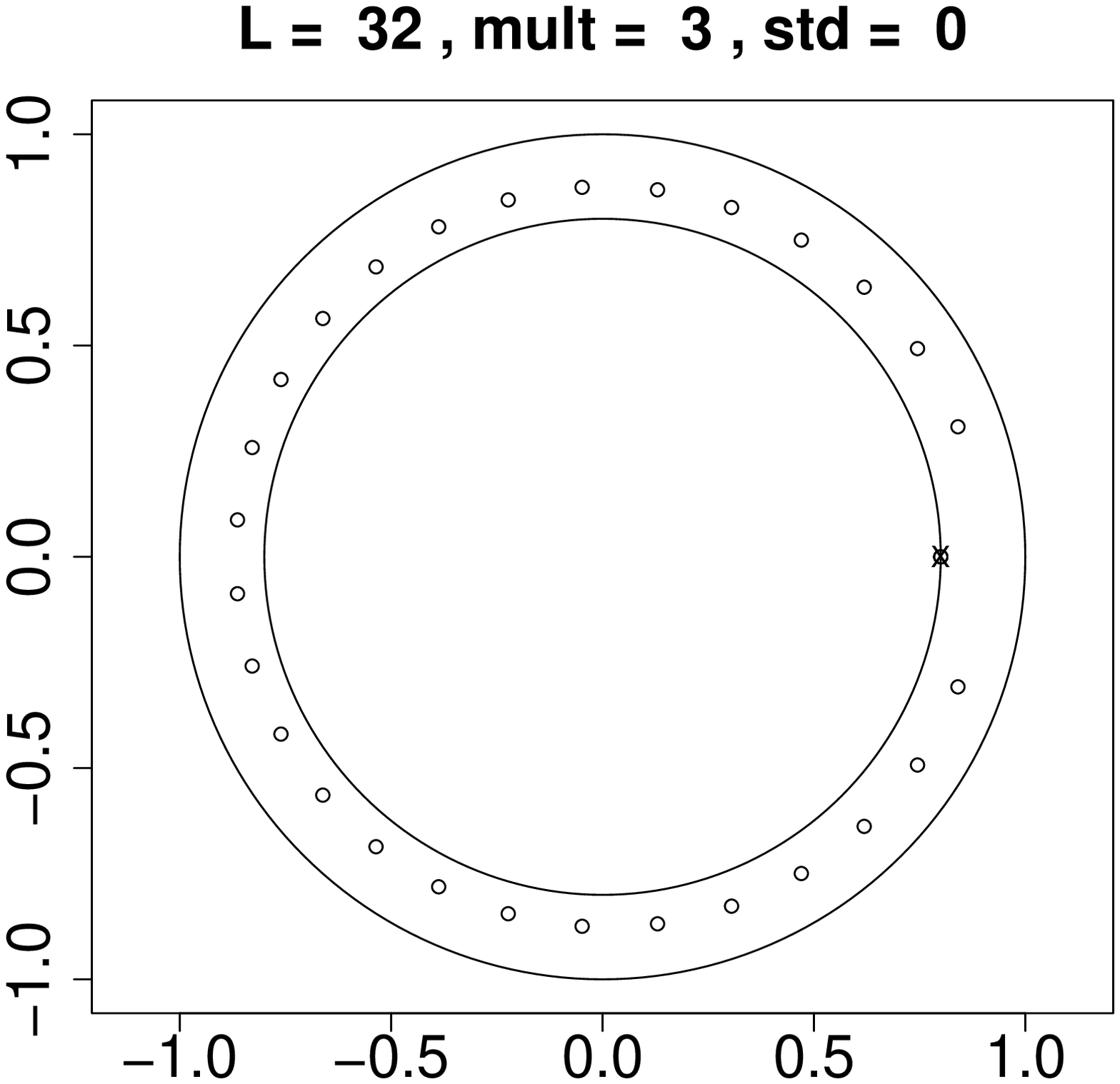}&\includegraphics[height=4.5cm]{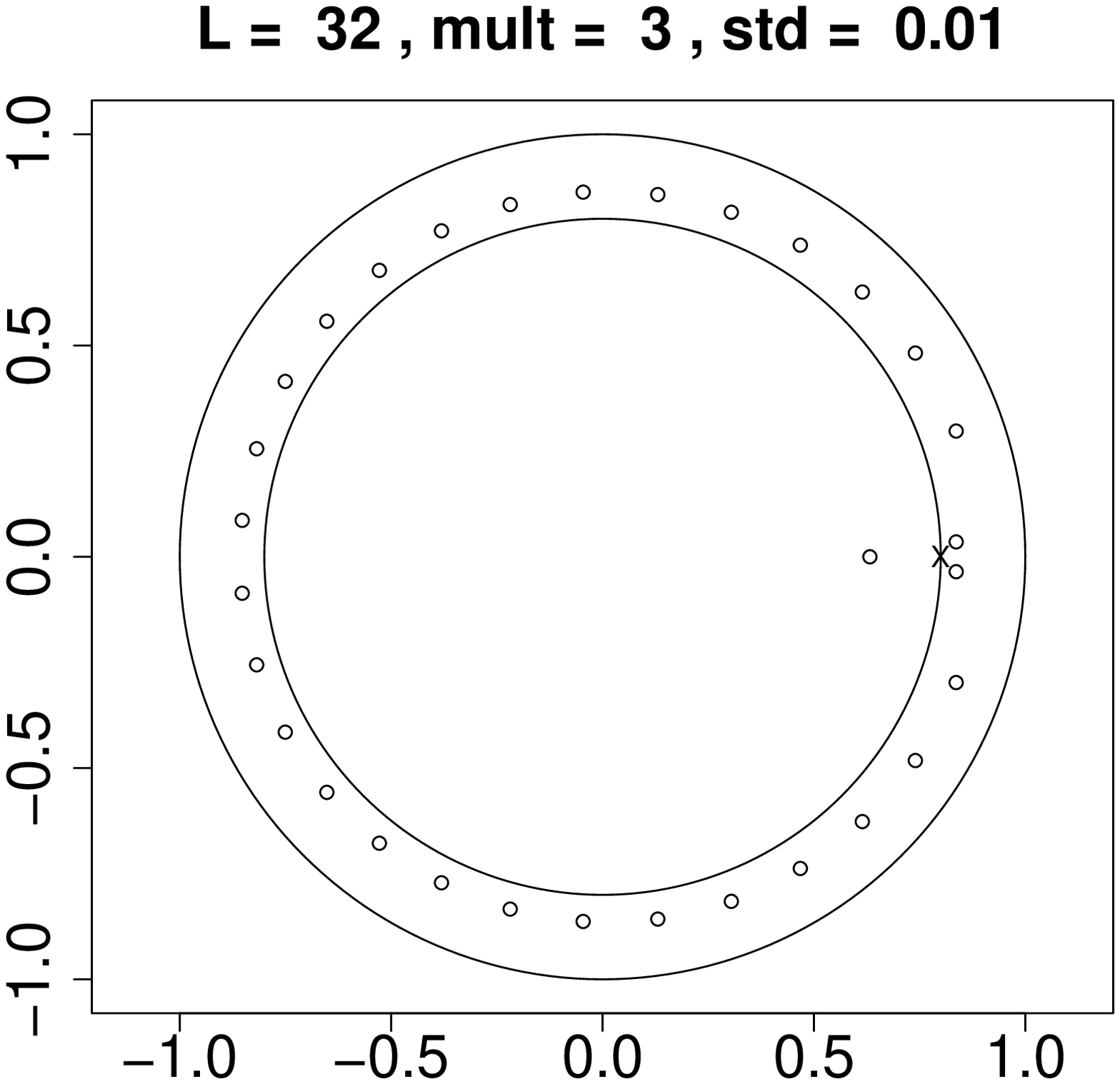}\end{array}
\eea
\caption{The signal roots (x) and the extraneous roots of the SSA LRF (o),
$f_n = n^2 0.8^n + {\rm std} \cdot \varepsilon_n$, $\varepsilon_n$ are i.i.d.  $\calN(0,1)$, $N = 150$.}
\label{fig:mult}
\efg
As one can see in Fig.~\ref{fig:mult}, in the presence of noise a multiple signal root splits into three separate roots.
Nevertheless, the extraneous roots behave as if there was a multiple signal root. Indeed, the extraneous roots are close to
those in the noise-free case. In addition, no ``spurious'' roots appear inside the critical circle.

\end{document}